\documentclass[12pt]{article}
\usepackage{amsmath,amssymb,graphicx} 
\usepackage{epsf}
\usepackage{subfigure}
\usepackage{cite}

\newcommand{\eg}{{\it e.g.}}

\def\beq{\begin{equation}}
\def\eeq{\end{equation}}
\def\beeq{\begin{eqnarray}}
\def\eeeq{\end{eqnarray}}
\def\be{\begin{eqnarray}}
\def\ee{\end{eqnarray}}
\def\beal{\begin{align}}
\def\eeal{\end{align}}

\begin{document}
\begin{titlepage}

\begin{flushright}
YITP-SB-10-21
\end{flushright}

\vskip.5cm
\begin{center}
{\large \bf  Template Overlap Method 
for Massive Jets}
\vskip.2cm
\end{center}

\begin{center}
 {Leandro G. Almeida}$^{a,b}$, {Seung J. Lee}$^c$, {Gilad Perez}$^c$, {George Sterman}$^a$, {Ilmo Sung}$^a$ \\

\vskip 8pt

$^{a}$ {\it \small C.N. Yang Institute for Theoretical Physics\\
Stony Brook University, Stony Brook, New York 11794-3840, USA}\\

\vspace*{0.3cm}

$^b$ {\it \small Physics Department, Brookhaven National Laboratory\\
Upton, New York 11973, USA}\\

   \vspace*{0.3cm}

$^c$ {\it \small Department of Particle Physics and Astrophysics \\
   Weizmann Institute of Science, Rehovot 76100, Israel}

\vspace*{0.853cm}

{\bf Abstract \vspace*{-0.1cm}\\}\end{center}

We introduce a new class of infrared safe jet 
observables, which we refer to as template overlaps,
designed to filter targeted highly boosted particle decays from QCD jets and other background.
Template overlaps are functional measures that quantify how
well the energy flow of a physical jet matches the flow of
a boosted partonic decay. 
Any region of the partonic phase space for the boosted decays defines a template.
We will refer to the maximum functional overlap found
this way as the template overlap.   To illustrate the method, we
test lowest-order templates designed to distinguish
highly-boosted top and Higgs decays  from backgrounds produced by event generators.
For the functional overlap, we find good results with a simple construction
based on a Gaussian in energy differences within angular regions surrounding the template partons.
 Although different event generators give 
 different averages for our template overlaps, we find in each case excellent rejection power,
 especially when  combined with cuts based on jet shapes.
 The template overlaps are capable of systematic improvement by including higher order
corrections in the template phase space.

\end{titlepage}
\newpage

\section{Introduction}
\label{intro} \setcounter{equation}{0} \setcounter{footnote}{3}

At the Large Hadron Collider, QCD will produce hadronic final states of unprecedented complexity,
and most searches for beyond-standard model physics will have to contend with large backgrounds. 
Over the past few years, scenarios have been proposed in which heavy particles,
including the Higgs and top quark, are produced at large transverse momentum~\cite{Butterworthsparticles,Butterworth:2002tt,Butterworth:2008iy,Butterworth:2009qa,KKG1,KKG2,Fitzpatrick:2007qr,Agashe:2007zd,Lillie:2007ve,Agashe:2007ki,Nath:2010zj}.  At high enough $p_T$, 
their decay products will appear as heavy, collimated jets
\cite{tscet,Banfi:2007gu}.   Even such exotic final states, however,
will coexist with a substantial tail of the mass distribution of light-parton QCD jets
\cite{Ellis:2007ib,Almeida:2008tp}, and it will generally be necessary
to study jet substructure systematically to distinguish such a signal.   

A number of methods
to analyze high-$p_T$ jets have been proposed and tested (so far) against the outputs of event
generators.   Generally, these methods depend on differences in the substructure of light-parton QCD jets
compared to those from particle decays.
Diagnostics to detect this
difference include infrared safe event shapes \cite{Thaler:2008ju,Almeida:2008yp}, and direct analyses of 
jet substructure~\cite{Butterworth:2008iy,Kaplan:2008ie,Pruning,Krohn:2009th,Kribs:2009yh,Chekanov:2010vc,Soper:2010xk}.
To this crowded field we propose a new method, based on a direct quantitative comparison of the energy flow of observed
jets at high-$p_T$ with the flow from specific partonic decay modes of boosted heavy particles.
Especially when combined with event shape information, the analysis of energy flow provides a potentially powerful tool.

Before going into details, we note that energy flow is a natural language for the description of 
jet structure.   Jet cross sections are naturally described in terms of correlation
functions of energy flow~\cite{Basham:1978bw}, which can be interpreted as correlations of
the energy-momentum tensor on the sphere ``at infinity"~\cite{Sveshnikov:1995vi,Korchemsky:1997sy,Bauer:2008dt,Hofman:2008ar}.
For QCD, these correlations tend to be strongly peaked, of course, around jets
that may represent the scattering or production of the partons of QCD or the decays
of short-lived resonances reflecting new dynamics.  

It is interesting to draw a contrast between QCD and the analogous problem for 
the cosmic background temperature, where the power distribution is very smooth.   Indeed,
motivated by observation as well as the inflation paradigm one expects for
this case a nearly scale invariant, almost featureless,
differential power spectrum. Hence, the CMB power spectrum, as well as the microscopic physics of the primordial epoch of inflation, is conveniently described by two and three point correlation functions of the power spectrum in momentum space~\cite{Maldacena:2002vr}.
Similarly, in case of conformal dynamics, the energy 
distribution resulting from hard scatterings
 can be well described by energy-energy correlation functions
  in momentum space~\cite{Hofman:2008ar}, and again is found to be smoothly distributed, almost spherically symmetric.
At first sight, energy flow in jet events could not be more different.   The search for the origin of
a given jet, however, whether from QCD radiation or from decay, may benefit from taking
a similar viewpoint, based on the pattern of energy correlations within jets.   In this paper, we will
present a method for such a quantitative study, with the aim of identifying jets that correspond to
resonance decay.   We will refer to this as a ``template" method, in which we use our knowledge
of the signal to design a custom analysis for each resonance, to make
use of differences  in energy flow between signal and background.

We can summarize the template overlap procedure as follows.
We denote by $|j\rangle$ the set of particles or calorimeter towers that
 make up a jet, identified by some algorithm, and take $|f\rangle$
to represent a set of partonic momenta $p_1\dots p_n$ that represent a boosted decay,
 found by the same algorithm.
We will introduce a functional measure ${\cal F}(j,f)\equiv \langle j | f \rangle$ that quantifies how
well the energy flow of $|j\rangle$ matches $|f\rangle$. 
Any region of partonic phase space for the boosted decays, $\{f\}$, defines a template.
We will often define the template overlap of observed jet $j$ as
$Ov(j,f)={\rm max}_{\,\{f\}}\, {\cal F}(j,f)$,
the maximum functional overlap of $j$ to a state $f[j]$ within the template region.
Template overlaps provide us with a tool to match unequivocally
arbitrary final states $j$ to partonic partners $f[j]$ at any 
given order.  
Once a ``peak template" $f[j]$ is found, we can use it to characterize the energy
flow of the state, which gives additional information on the likelihood that
it is signal or background.

To make the matching between physical and template
possible, each event is characterized by some set of particle or calorimeter energies,
$E(\theta_i,\phi_j)_{i,j\in R}$, where $E$ is the energy and
$\theta$ and $\phi$ represent coordinates internal to a jet
with cone or related parameter $R$.   In a typical experimental setup the energy is discretized according to the detector resolution, and each pair $i,j$ corresponds to a specific cell in the calorimeter.
At the LHC experiments~\cite{TDRs}, for instance, electromagnetic calorimeter cell size (in $\eta$ and $\phi$) is of ${\cal O}\big(0.025\times0.025\big)$ and of ${\cal O}\big(0.1\times0.1\big)$ for hadronic calorimeter cells.
For each event, the overlap with the template states is calculated.  

In general, for each state $j$, the template state $f[j]$ with  maximal overlap with $j$ will be used to characterize the event $j$.
We therefore adopt the ansatz that a good (if not the best)  rejection power is obtained when we use the signal 
distribution itself  to construct our templates (see \eg~\cite{Maldacena:2002vr}).  At lowest order all the information encoded in the events
is matched uniquely to the lowest order template with maximum overlap.   After showering and hadronization,  this correspondence
is diluted, but as we shall see, very meaningful correlations remain.
  
The application of these ideas is particularly straightforward for top jets.  Much of
the QCD background is characterized by two sub-jets, 
with very different energy flow from the three-parton templates in general.
Indeed, for a lowest order partonic QCD jet consisting of the original parton
plus one soft gluon, there is no template state from top decay that 
matches the energy flow.   This gives a fundamental discrimination, to which
we can add additional information from event shapes.  
 
 Having given a rationale for the template method,   in the following section we 
 provide a general formalism to describe it. In Sec.\ \ref{3point} we apply the method 
 to templates tailored to a boosted top search.  
 In this case, as noted above,
 the three-particle structure of the lowest-order templates gives a clear
 distinction between signal and background, which we amplify further by the
 use of other infrared safe event shapes.   Comparisons are carried out
 using anti-$k_T$ jet finders for events found from several Monte Carlo (MC) generators.
 In each case, we find large background rejection powers based on this analysis, with
 substantial efficiencies.   
 
  Highly boosted Higgs decays are discussed in Sec.\ \ref{2point}.   In this case,
  the signal and background are both two-parton states at lowest order~(LO).    Their template overlap
  distributions are slightly different, but here we use another feature of
  the template method: the uniqueness of the template state with maximum overlap.
 This information provides us with an additional, infrared safe
  tool, which will enable us to attain significant rejection power even in this case.   
  We conclude in Sec.\ \ref{conclusion}. 

\section{Overlap Formalism}\label{general} 

We want our template overlaps to be functionals of energy flow of any specific
event (usually involving jets), which we label $j$, and a model, or template,
for the energy flow in a signal, referred to as $f$ .
Our templates will be a set of partonic momenta $f=p_1\dots p_n$, with 
\beeq
\sum_{i=1}^n p_i=P\, ,\, \quad P^2=M^2\, , 
\eeeq
which we take to represent the decay products of a signal of mass $M$.
For example, the lowest-order template for Higgs decay would have
$n=2$ and for top decay, $n=3$.   
Of course, templates with more than the minimum number of particles are
possible.   
To represent the sum over this $n$-particle phase space, we
introduce the notation
\beeq
\tau^{(R)}_n \equiv
\int \prod_{i=1}^n \frac{d^3\vec p_i}{(2\pi)^3 2\omega_i}\ \delta^4(P-\sum_{i=1}^n p_i)\ \Theta(\{p_i\},R)\, ,
\eeeq
where the function $\Theta(\{p_i\},R)$ limits the phase space integral to some
region, $R$, which may represent a specific cone size, for example.

We would like to measure how well the energy flow of any given event $j$
matches that of the signal on the unit sphere, denoted
by $\Omega$.   We represent the template energy flow as
$d E{(f=p_1\dots p_n)}/d \Omega$.   This function is taken at 
fixed (to start with, lowest) order.   Similarly, we will represent the 
energy flow of event $j$ as $d E{(j)}/d \Omega$.   This quantity is observed, either
in experiment or the output of an event generator.
Schematically, a general overlap functional $Ov(j,f)$ is
represented as  
\beq
Ov(j,f) = \langle j|f\rangle = {\cal{F}}\left[ \frac{d E{(j)}}{d \Omega} , \frac{d E{(f)}}{d \Omega}  \right]\, .
\eeq
In principle, the choice of the functional $\cal F$ is arbitrary.   

A natural measure  of the matching
between state $j$ and the template is the weighted
difference of their energy flows integrated over some specific region
that includes the template momenta $p_i$.  
To quantify this difference, we construct the functional $\cal F$ using the
template states.
We will find it useful to identify the difference in terms of the template 
configuration in $n$-particle phase space with the
{\it closest match} of energy flow to a given state $j$.   
As a measure of the matching we introduce a function $\Phi(x)$
that is maximized at $x=0$ to $\Phi(0)=1$, which represents a ``perfect" match.
A simple example, which we will employ below,
 is a Gaussian,
\beeq
Ov^{(F)}(j,f)  
={ \rm{max}}_{\tau^{(R)}_n}\ \exp\left[\, -\, \frac{1}{2\sigma^2_E}\left(  \int d \Omega\,\,  
 \left[  \frac{d E{(j)}}{d \Omega}  - \frac{d E{(f)}}{d \Omega}  \right ]
  F ( \Omega,f) \right)^2 \right]\, ,
  \label{overlap1}
\eeeq
where we introduce a width, $\sigma_E$ with units of energy.
For infrared safety, the function $F(\Omega,f)$ should be a sufficiently smooth
function of the angles for any template state $f$\ \cite{Sterman:1979uw}.
For example, it could be defined as a Gaussian around each of
the directions of the template momenta \cite{Lai:2008zp}.
Alternately, we may choose $F$ to be a normalized  step function
that is nonzero only in definite angular regions around the directions of the
template momenta $p_i$\ \cite{Sterman:1978bj}.
This is the method we will use below.
We emphasize that the choice of our overlap functional is 
to a large extent arbitrary, subject to the requirements of infrared safety.
We will find, however, that relatively simple choices can give
strong enrichment of signals.

To be specific, for an $n$-particle final state, we will represent our
template overlap (dropping the superscript $(F)$) as 
\beeq
Ov(j,p_1\dots p_n)  &=& 
{ \rm{max}}_{\tau^{(R)}_n}\ 
\exp\left[\, -\ \sum_{a=1}^n \frac{1}{2\sigma^2_a}\left(  \int d^2 \hat n\,
  \frac{d E{(j)}}{d^2 \hat n} \theta ( \hat n, \hat n_a^{(f)})  - E_a^{(f)}\, \right)^2
   \right]\, ,
  \label{overlap2}
\eeeq
where the direction of template particle $a$ is $\hat n_a$
and its energy is $E_a^{(f)}$.   In applications below, we will
use these energies to set the widths of the Gaussians.
The functions $ \theta (\hat n, \hat n_a^{(f)})$
restrict the angular integrals to (nonintersecting) regions
surrounding each of the template momenta.
We will refer to the corresponding state as the ``peak template" $f[j]$ for state $j$.   
The peak template $f[j]$ provides us with potentially valuable information
on energy flow in $j$.   

In summary,  the output of 
the peak template method for any physical state $j$ is 
the value of the overlap, $Ov(j,f)$, and also the 
identity of the template state $f[j]$ to which the best match is found.
As we shall see, this will be of particular
value when we apply our method to boosted Higgs.   We turn first,
however, to the analysis for boosted tops.

\section{Three-particle Templates and Top Decay}\label{3point}

In this section, we illustrate the peak template method for top identification, using
as a template the LO partonic three-particle phase space of top decay.
The essential observation is that light-quark and gluon jets (generally referred to as ``QCD jets" below)
typically have states with a two-subjet topology.   Such states generally do not
match well with a three-particle template, and so are easy to separate from
the signal on the basis of their low values of $Ov$\cite{Almeida:2008yp,Almeida:2008tp}.   
 Of course, some 
top decay states have low values of $Ov$ also, and some QCD jets
higher values.   We will see how to combine the template overlap
with planar flow to develop a filter that enriches the top signal at
relatively high efficiency.

\subsection{Peak template overlap method}

We begin with a detailed description of the peak template overlap method 
with the LO three-parton templates appropriate to top jet analysis.

\subsubsection{Mass cut and discretization of data}

First, we select data using a jet mass window for the top, choosing  160 GeV $\le m_J \le$ 190 GeV, with top mass chosen to be 174 GeV\footnote{We choose this value for the purpose of demonstration only, and the running of the top mass 
may be important.}, cone size $R=0.5$ ($D=0.5$ for anti-$k_T$ jet algorithm
\cite{Cacciari:2008gp}
 we are using) and jet energy 950 GeV $\le P_0 \le$ 1050 GeV.
In our demonstration, we choose a discretized $\theta$-$\phi$ plane,
with $\Delta \theta=0.06$ and $\Delta \phi=0.1$.
Then, we can build a table of energy $E(\rm row_m, column_n)$, where $\rm row_m$ and $\rm column_n$ are the row and column number corresponding to the discretized $\theta$ and $\phi$.

\subsubsection{Construction of template states}\label{3point_construction}

We wish to generate a sufficient number of template states to cover three-particle phase space
for top decay, $t\rightarrow b+W\rightarrow b+q+\bar q$.   Imposing the condition, $(p_q+p_{\bar q})^2=M_W^2$,
there are four degrees of freedom.    To construct our set of states, we have chosen a brute force
method, based on four angles.   We take two of these to be the polar and azimuthal angles
that define the $b$ and $W$ directions in the top rest frame, defined relative to the direction of the
boost from this frame to the lab frame.   The remaining two are again polar and azimuthal angles,
that define the $q$ and $\bar q$ directions, this time relative to the boost axis from the $W$ rest frame
to the top rest frame.   This method is by no means unique.   

By straightforward Lorentz transformations of particle momenta,
the four angles identified above determine the energies and directions of the three decay products
of the top at LO.   We neglect the possible effects of spin and polarization at the particle level in our construction of template states.\footnote{In~\cite{Almeida:2008yp} it was shown that for boosted two pronged decays, energy flow is very similar for massive spin zero and spin one.}

For this investigation, we discretize all four physical angles with a discretization length of $0.1$. As for the discretization of the data, we encode two physical angles in terms of row and column number corresponding to the data discretization scheme.
A given template consists of a list ($\rm row_a, column_a$, $E_{\rm a}$, a=1,2,3) for each of three daughter particles of hardronic top ($b$, $q$ and $\bar q$).  We exclude those templates having particles whose polar angles, $\theta$ 
relative to the jet axis, are larger than the cone size $R$. Also, we impose an energy cut on the templates, removing templates that have an energy less than 10 percent of the maximum energy for a given particle.
The number of template states, constructed as above,
that pass these cuts is very large, of order three million.   We are confident, therefore, that the
maximum overlap found with this set is very close to the true maximum.   We emphasize that, once
generated, the same set of template states is used for all the data.

\subsubsection{The template overlap}\label{3point_convolution}

We next define an overlap between template, $|f\rangle $, and
a specific jet energy configuration $|j\rangle$,  $\langle j | f \rangle$.   Following Eq.\ (\ref{overlap2}), we set
\beeq
Ov(j,f)  
=
{ \rm{max}}_{\tau^{(R)}_n}\ 
\exp\left[\, -\ \sum_{a=1}^3 \frac{1}{2\sigma^2_a}\left(  
\,  \sum_{k=i_a-1}^{i_a+1}\sum_{l=j_a-1}^{j_a+1} E{(k,l)}   - E(i_a,j_a)^{(f)}\, \right)^2
   \right]\, ,
\label{3particletemplate}
\eeeq
where $E(i_a,j_a)^{(f)}$ is the energy for the template particle $a$, 
whose direction is labelled by indices $i_a$ and $j_a$, according to the discretization table
described above.
For our analysis, we fix $\sigma_a$ (for the $a$th parton) by that parton's energy,
\beq
\sigma_a=E(i_a,j_a)^{(f)}/2\, .
\label{overlap_convolution2}
\eeq 
In Eq.\ (\ref{3particletemplate}), we define the overlap between data state $j$ and template $f$
on the basis of an unweighted sum of all the energy in the total of nine cells 
of state $j$ surrounding (and including) each of the three cells populated by a particle in state $f$.   
If one of the cells is located on the edge of the cone in the 
 direction of the polar angle with respect to the jet direction, the number of cells included
 in the sum is simply taken to be smaller. 

\subsection{Peak template overlaps for top and QCD jets}

We can now apply the peak template function method discussed in the previous 
sections to analyze energetic top jet events {\it vis-a-vis} QCD jets. We use the data
for QCD jet and hadronic top jet events, for $R=0.5$, 950 GeV$\le P_{0} \le$1050 GeV, 160 GeV$\le m_J \le$190 GeV and  $m_{top}=174$ GeV as obtained
via the anti-$k_T$ jet clustering algorithm~\cite{Cacciari:2008gp} with CTEQ6M PDF set~\cite{Pumplin:2002vw}.  
The main purpose of this section is to understand how well we can discriminate our signal from the potentially overwhelming
QCD background by using the simplest three-point correlation template functions. 

\begin{figure}[hptb]
\begin{tabular}{ccc}
\includegraphics[width=.48\hsize]{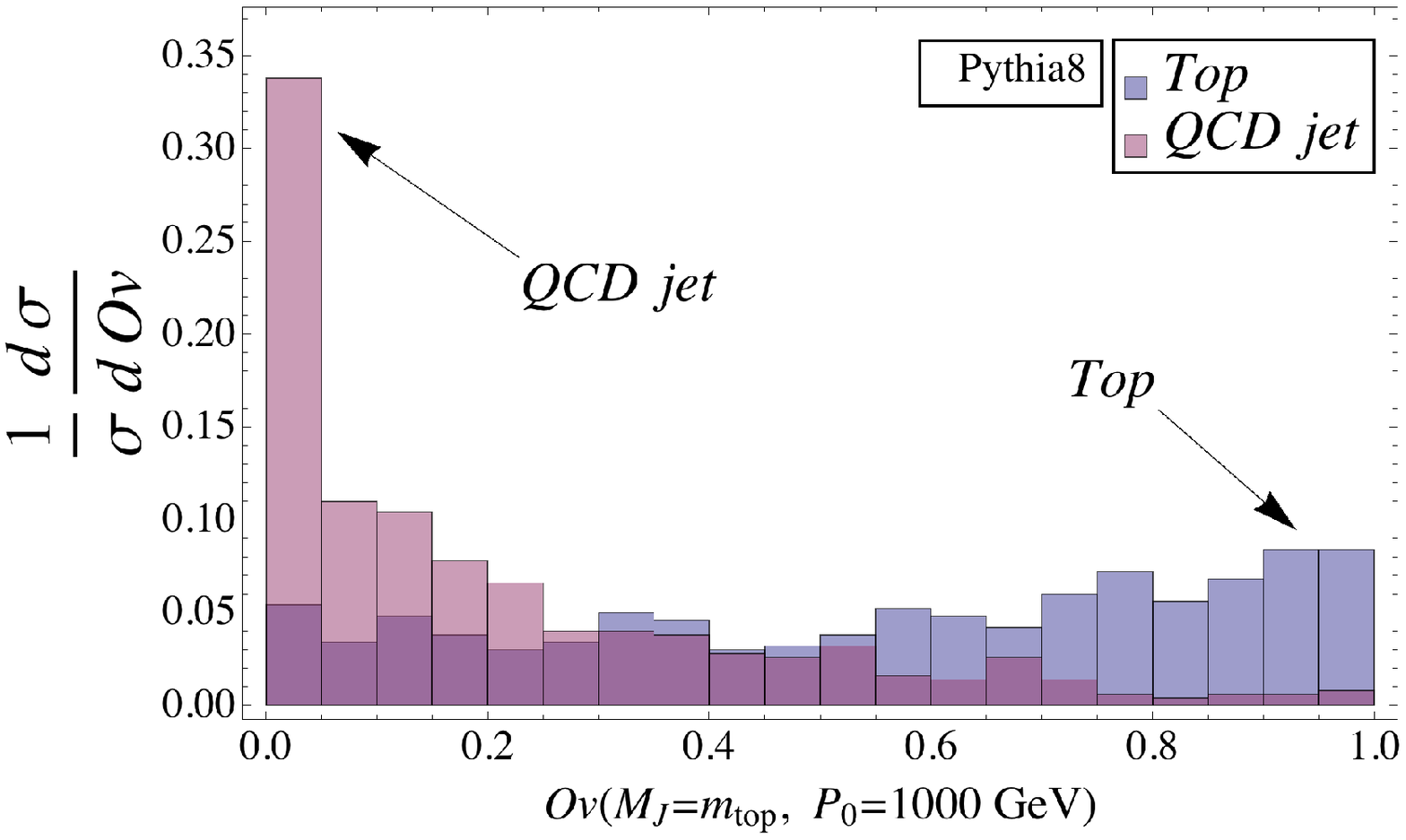}
\includegraphics[width=.48\hsize]{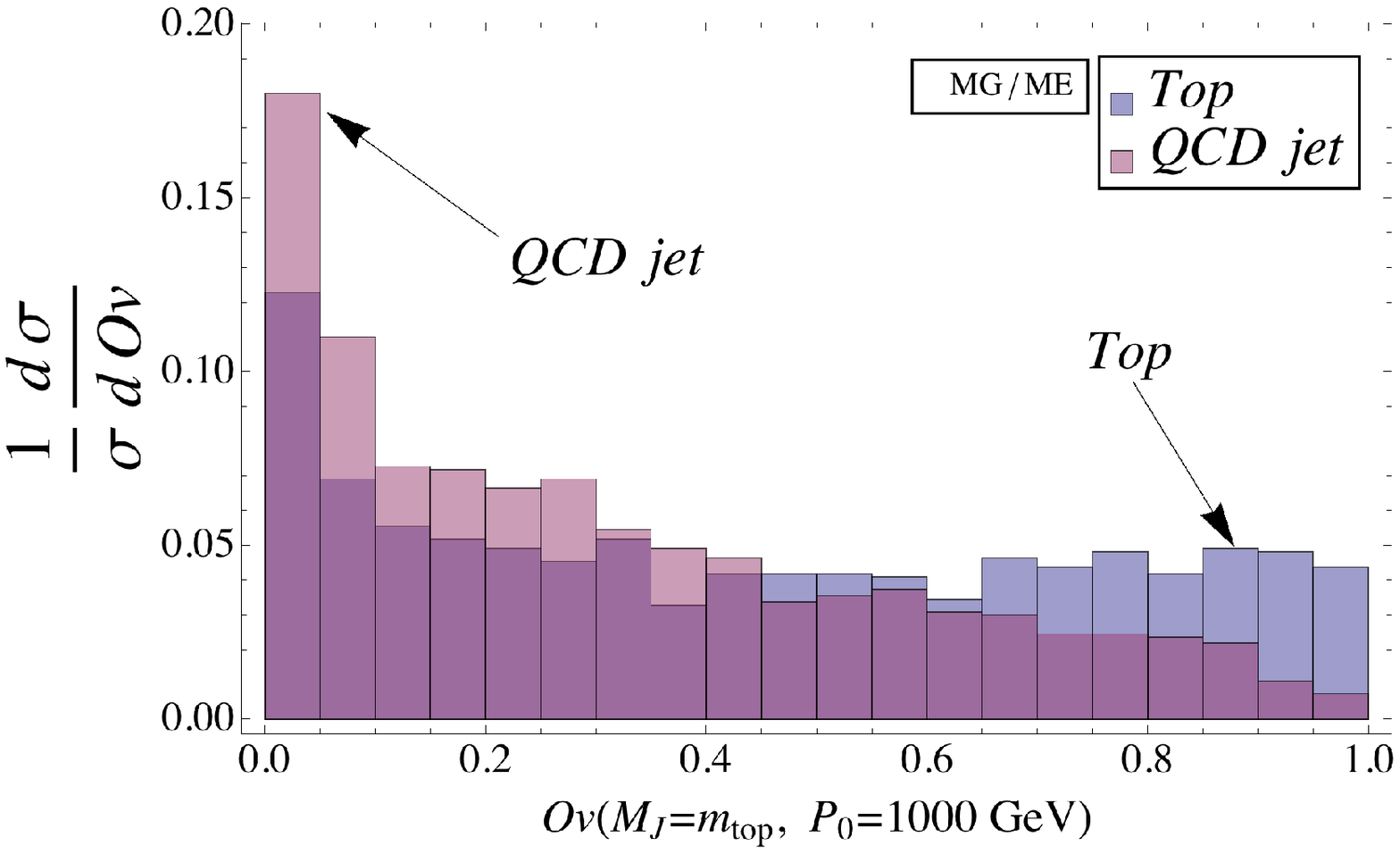} \\
\includegraphics[width=.48\hsize]{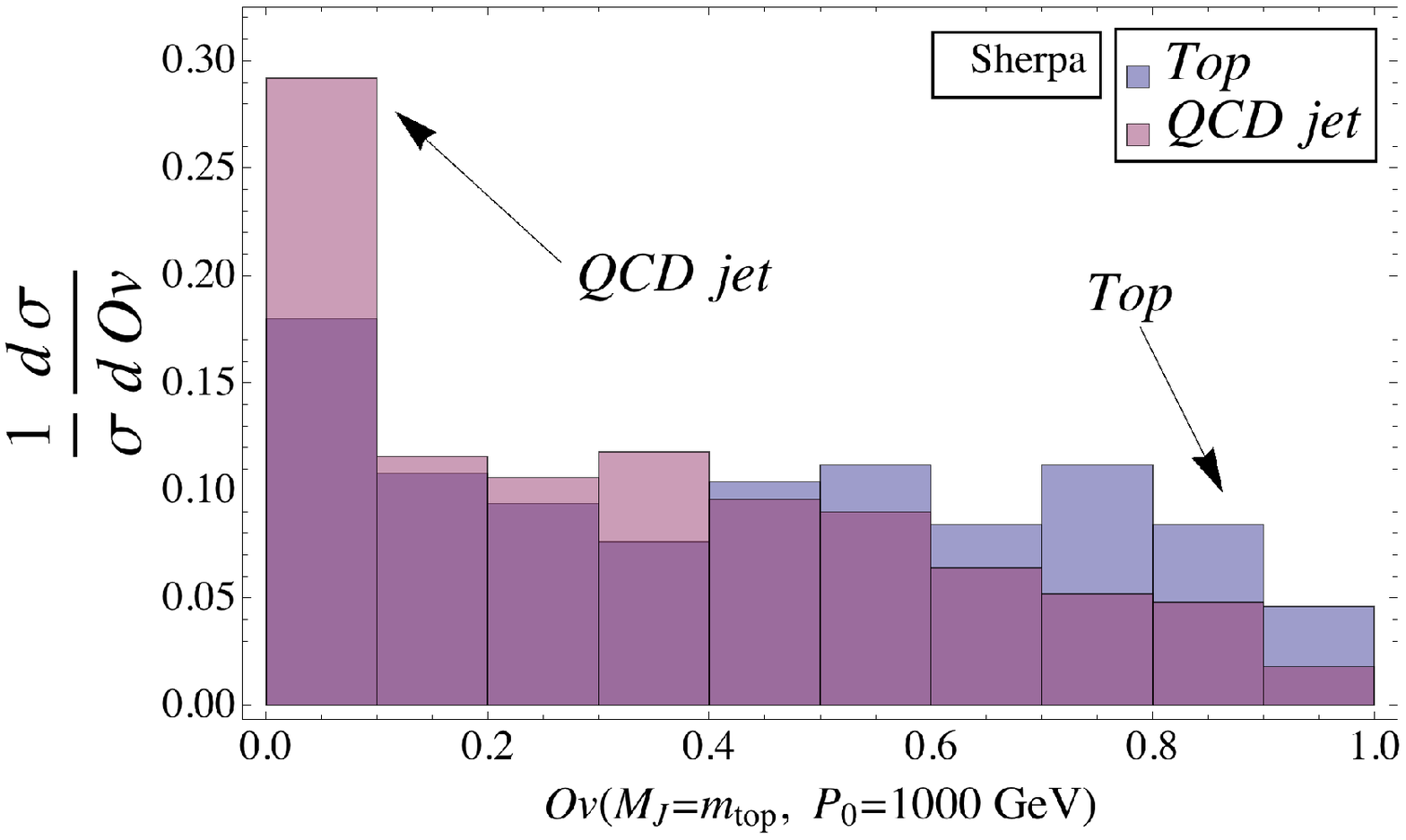}   
\end{tabular}
\caption{Comparison of histograms of template overlap $Ov$, Eq.~(\ref{3particletemplate}), with top jets and QCD jets from different MCs [upper left (right) Pythia  (MG/ME) and Sherpa on the bottom], for $R=0.5$,  950 GeV$\le P_{0} \le$1050 GeV, 160 GeV$\le m_J \le$190 GeV and  $m_{top}=174$ GeV.
}\label{top_probe_function_comparison}
\end{figure}

In Fig.~\ref{top_probe_function_comparison} we compare the overlap distributions for showered top jets and QCD jets (for the same $z=m_J/P_0$) for event generators
Pythia (version 8)~\cite{Sjostrand:2007gs} for $2\rightarrow 2$ process without matching,
MadGraph/MadEvent (MG/ME) 6.4~\cite{MG} (with MLM matching \cite{MLM} interfaced into Pythia V6.4~\cite{Sjostrand:2006za}), 
and Sherpa 1.2.1~\cite{Gleisberg:2008ta,Hoeche:2009rj}. It is clear that the showering smears the top distributions significantly, although top events tend to yield somewhat larger peak overlaps.   Note also the large variations between the generators.
 
\subsection{Planar flow}

We have seen that LO top templates already distinguish noticeably between
top and QCD jets.   There is still a close relation, with both distributions being fairly flat.  To gain a better resolution
between the two possibilities, we shall rely on the
 jet shape variable, planar flow~\cite{Almeida:2008yp,Thaler:2008ju}.
For completeness we give the definition of the planar flow variable, $Pf$.
First construct for a given jet, a matrix  
$I_{\omega}$  as
\be
I^{kl}_{\omega}= {1\over m_J} \sum_i {\omega}_i \frac{p_{i,k}}{{\omega}_i}\,\frac{p_{i,l}}{{\omega}_i}\, ,
\ee
where $m_J$ is the jet mass, ${\omega}_i$ is the energy of particle $i$ in the jet,
and $p_{i,k}$ is the $k^{th}$ component of its transverse momentum relative to the 
axis of the jet's momentum.
The $Pf$ variable is defined as
\be
Pf={4\,{\rm det}(I_{\omega})\over{\rm tr}(I_{\omega})^2}={4 \lambda_1 \lambda_2\over(\lambda_1 + \lambda_2)^2}\, ,
\label{Pfdef}
\ee
where $\lambda_{1,2}$ are the eigenvalues of $I_{\omega}$.
We shall see that planar flow distinguishes between many three-jet events with large
template overlaps.    In general, QCD events with large $Ov$ will have significantly smaller planar 
flow than top decay events.   
For the QCD jets a large overlap would be a result of a kinematic ``accident".
In the studies we show below, the combination of $Ov$ and $Pf$ gives a
strong background (QCD) suppression with quite substantial signal (top decay) efficiency.

\begin{figure}[hptb]
\begin{center}
\begin{tabular}{c}
\includegraphics[width=.40\hsize]{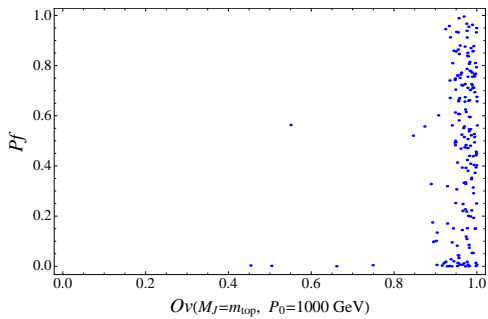} 
\end{tabular}
\caption{A scatter plot of template overlap, Eq.~(\ref{3particletemplate}) and $Pf$ for  LO parton-level MC output for top quark decay, with $P_{0}=1$ TeV, $m_{top}=174$ GeV. 
}\label{partonic_level_top}
\end{center}
\end{figure}

In Fig.~\ref{partonic_level_top}, we test these ideas by plotting the template overlap $Ov$  for the partonic level 
output of a MC, versus $Pf$.
The data shows a scatter plot of  $Ov$ and $Pf$ found in this way.   The
data are all close to unity in $Ov$, but are (as expected) spread out in planar flow.
As we may conclude by looking back at Fig.\ \ref{top_probe_function_comparison}, the effect of showering is to spread out top
decays over the full range of $Ov$.   

\subsection{Application to top decay}

\begin{figure}[hptb]
\begin{tabular}{ccc}
\includegraphics[width=.48\hsize]{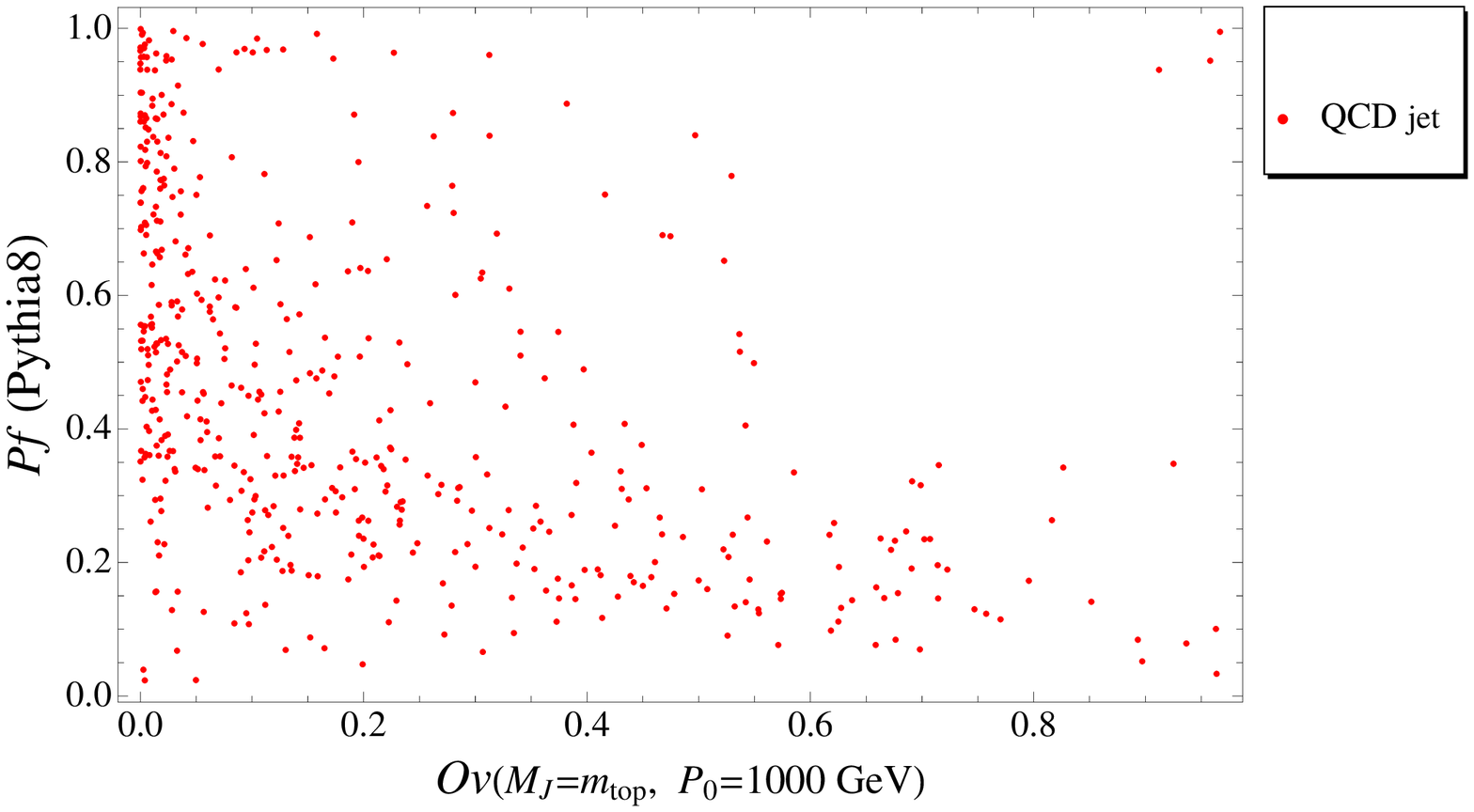}
\includegraphics[width=.48\hsize]{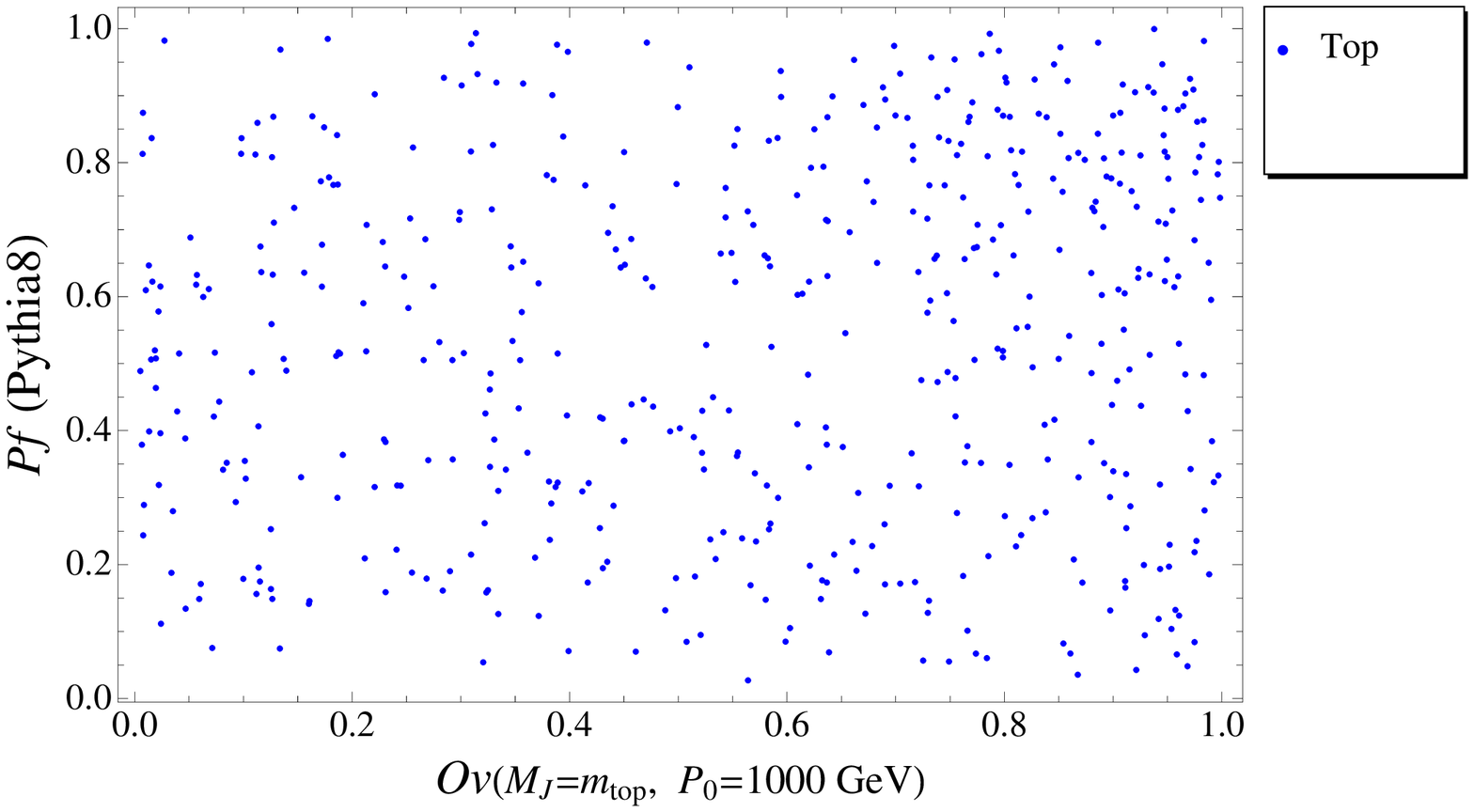} \\
\includegraphics[width=.48\hsize]{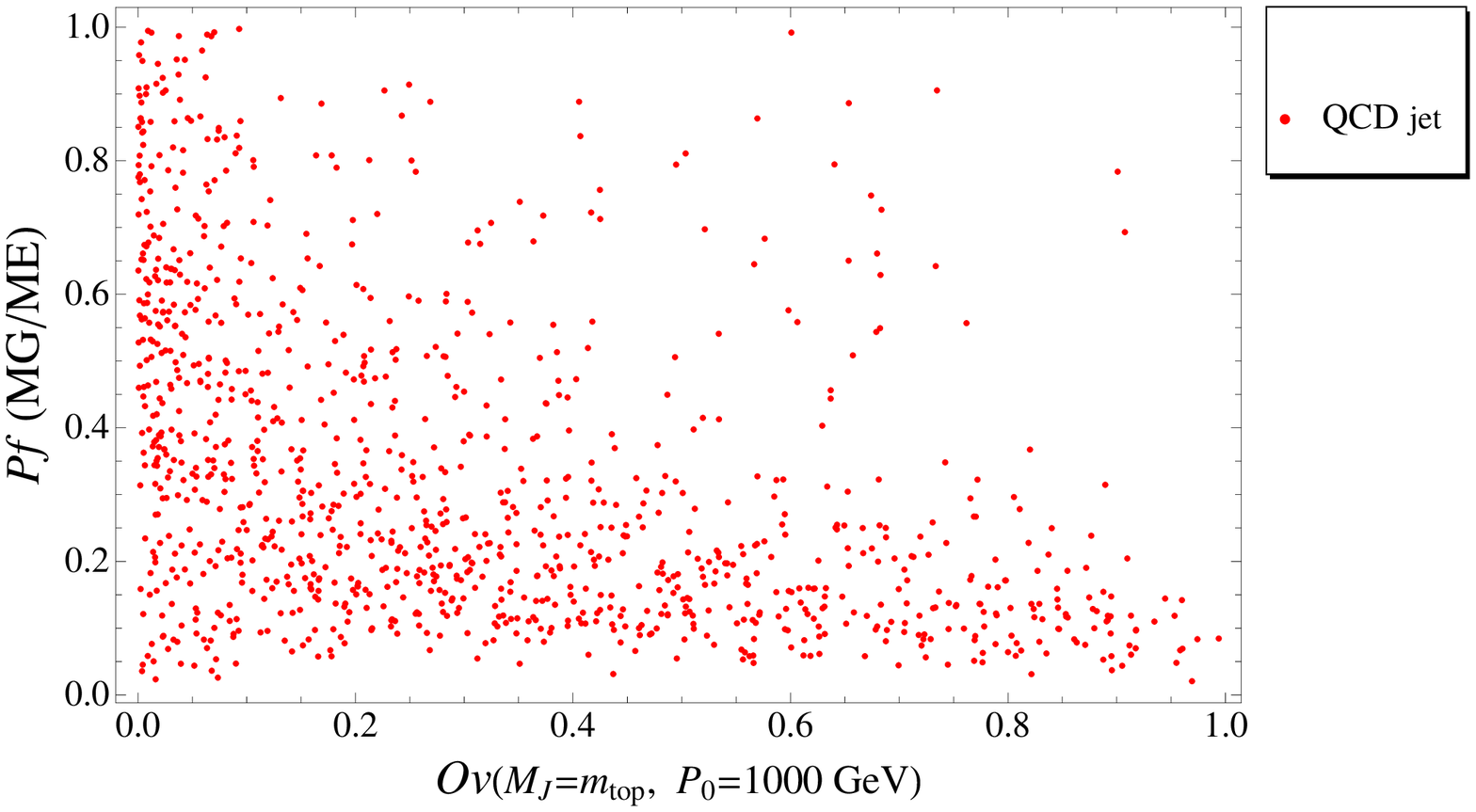}
\includegraphics[width=.48\hsize]{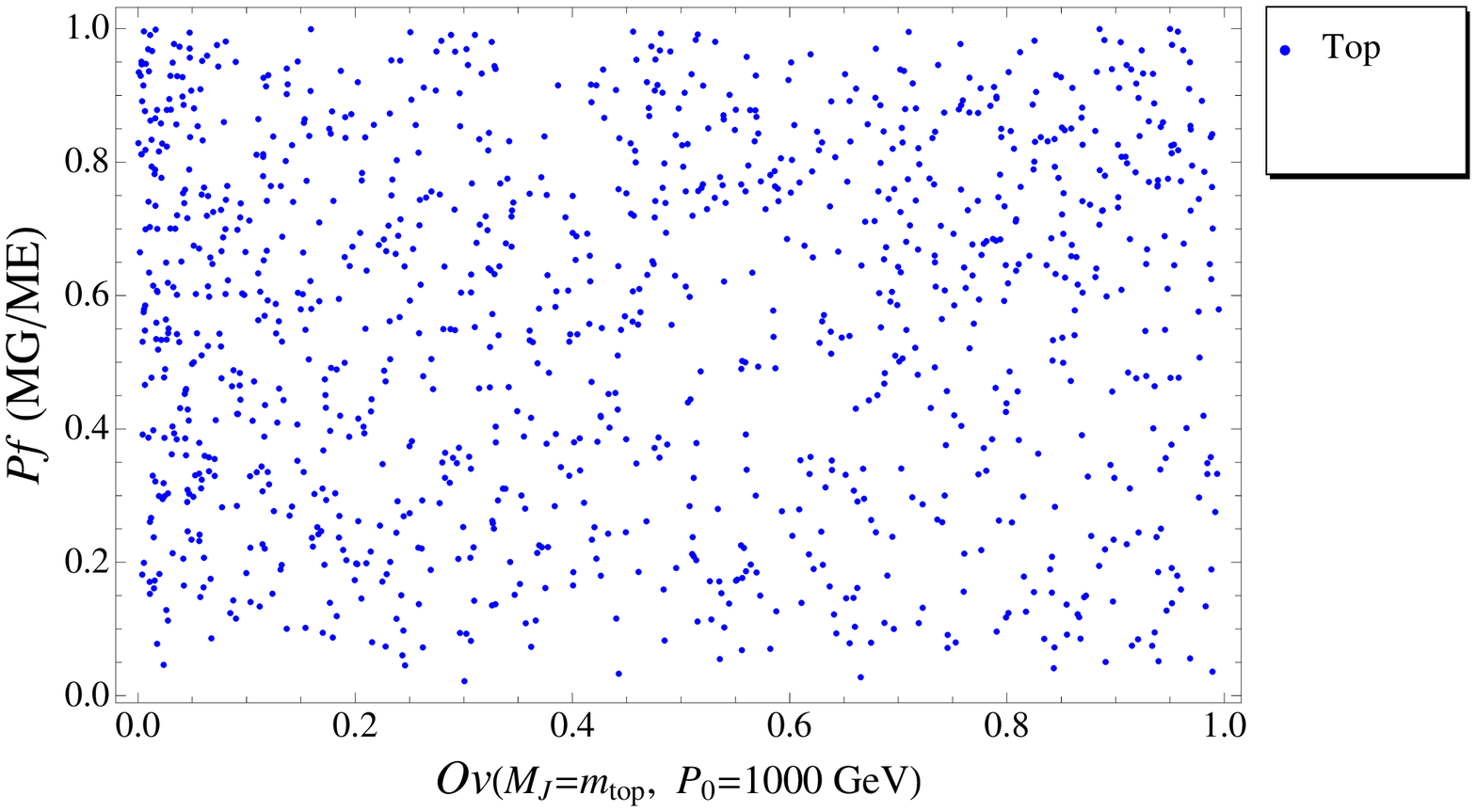} \\
\includegraphics[width=.48\hsize]{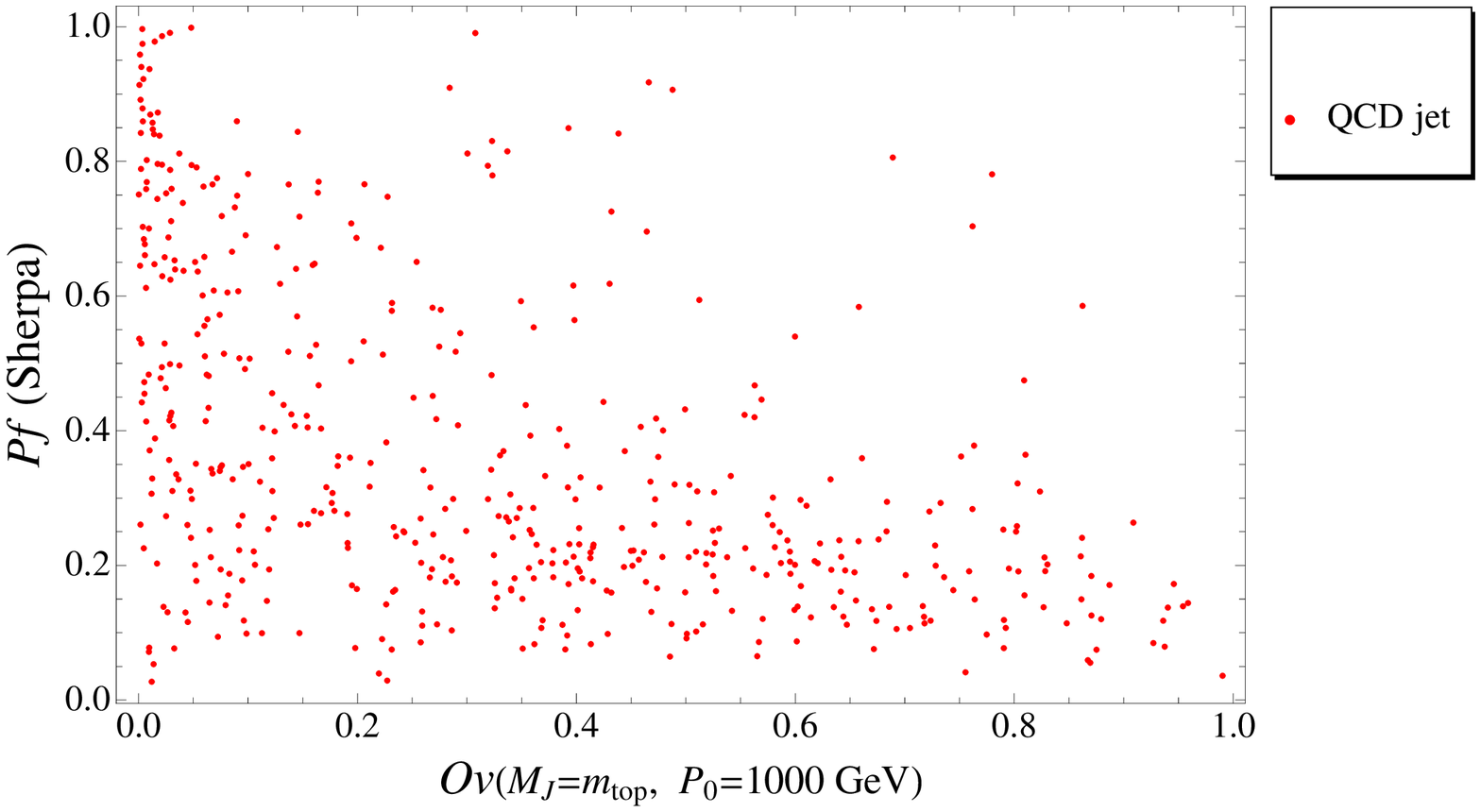}   
\includegraphics[width=.48\hsize]{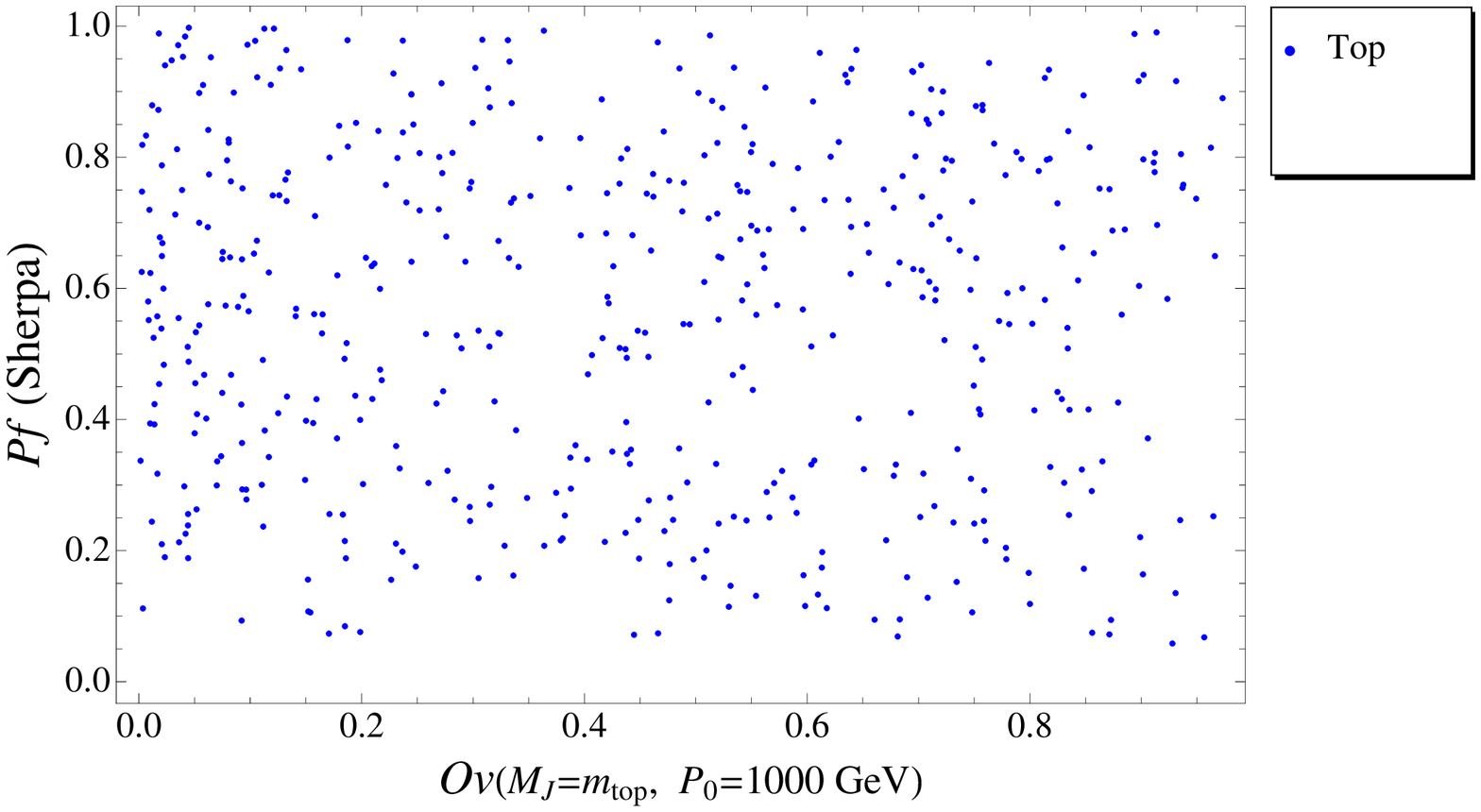}   \\

\end{tabular}
\caption{
Comparison of scatter plots of planar flow $Pf$  {\it vs.}\ template overlap $Ov$ for top jets (right) and QCD jets (left) from different MC (from top to bottom: Pythia, MG/ME, Sherpa), for $R=0.5$,  950 GeV$\le P_{0} \le$1050 GeV, 160 GeV$\le m_J \le$190 GeV and  $m_{top}=174$ GeV.
}\label{top_probe_function_pf_comparison}
\end{figure}

In Fig.\ \ref{top_probe_function_pf_comparison} we show a 
comparison of scatter plots of planar flow, $Pf$ {\it vs.}\ template overlap, $Ov$ with QCD (first column) top jets 
(second column) from different MC (from top to bottom: Pythia, MG/ME, Sherpa), for $R=0.5$,  950\ GeV$\le P_{0} \le$1050\ GeV, 160 GeV$\le m_J \le$190 GeV and  $m_{top}=174$\ GeV. The 
three event generators provide rather different distributions, but in
each case the distinction between the signal and background distributions is evident.
Clearly, any set of events 
chosen from the upper right of these plots, with $Pf > Ov$, is
highly enriched in top events compared with background.   
The clear differences in these scatter plots show the potential of the template overlap method.

\begin{figure}[hptb]
\begin{tabular}{ccc}
\includegraphics[width=.5\hsize]{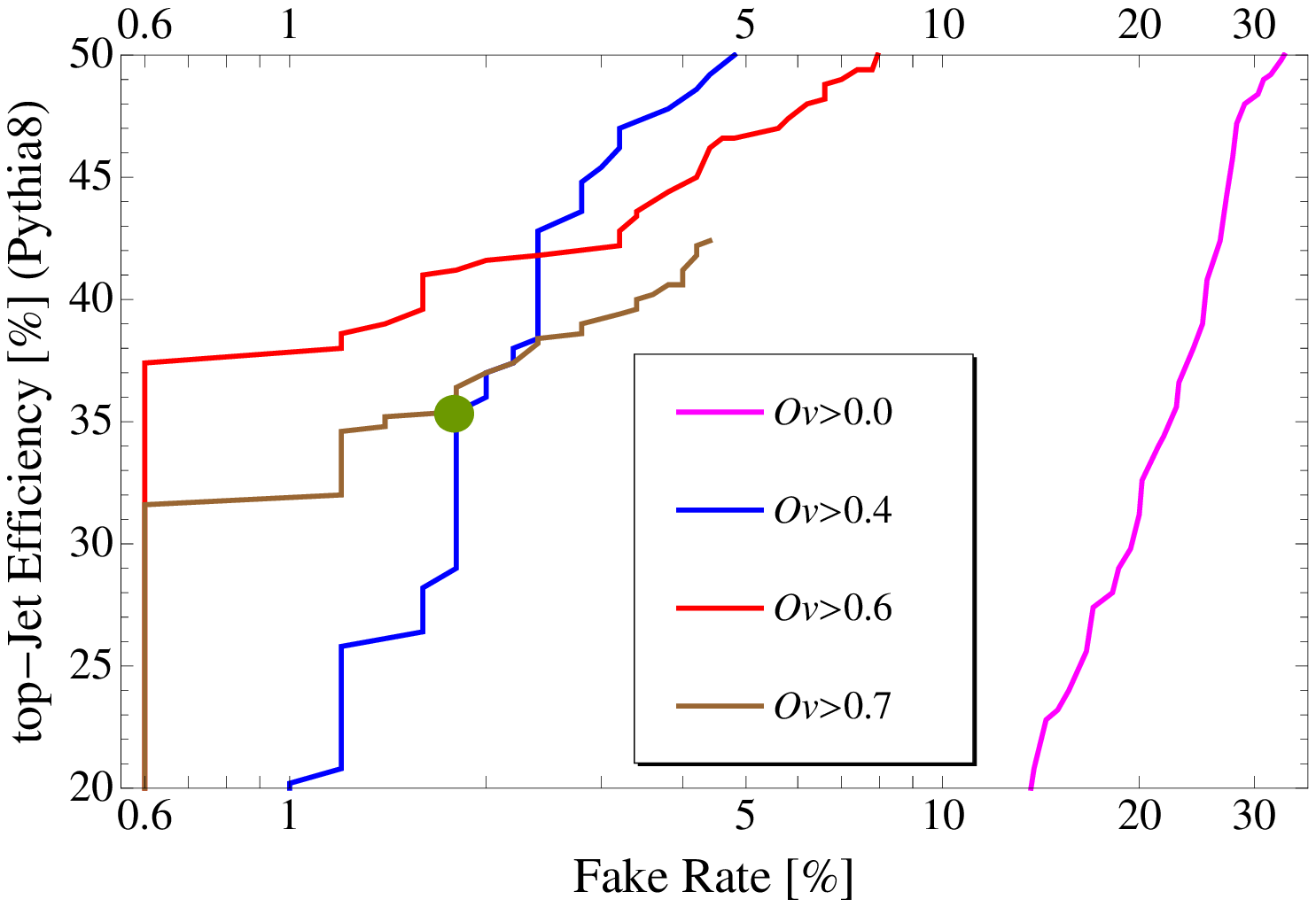}
\includegraphics[width=.5\hsize]{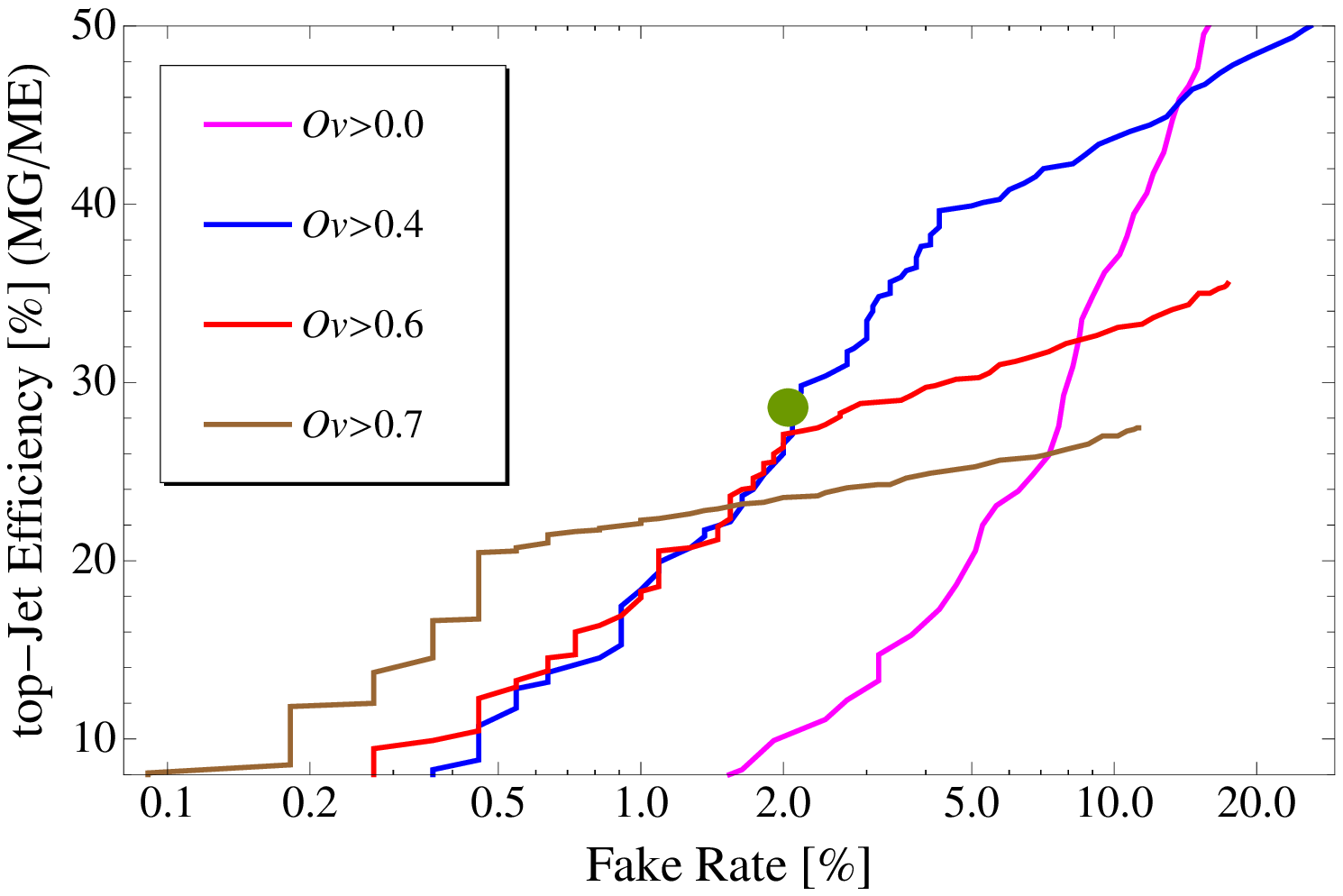} \\
\includegraphics[width=.5\hsize]{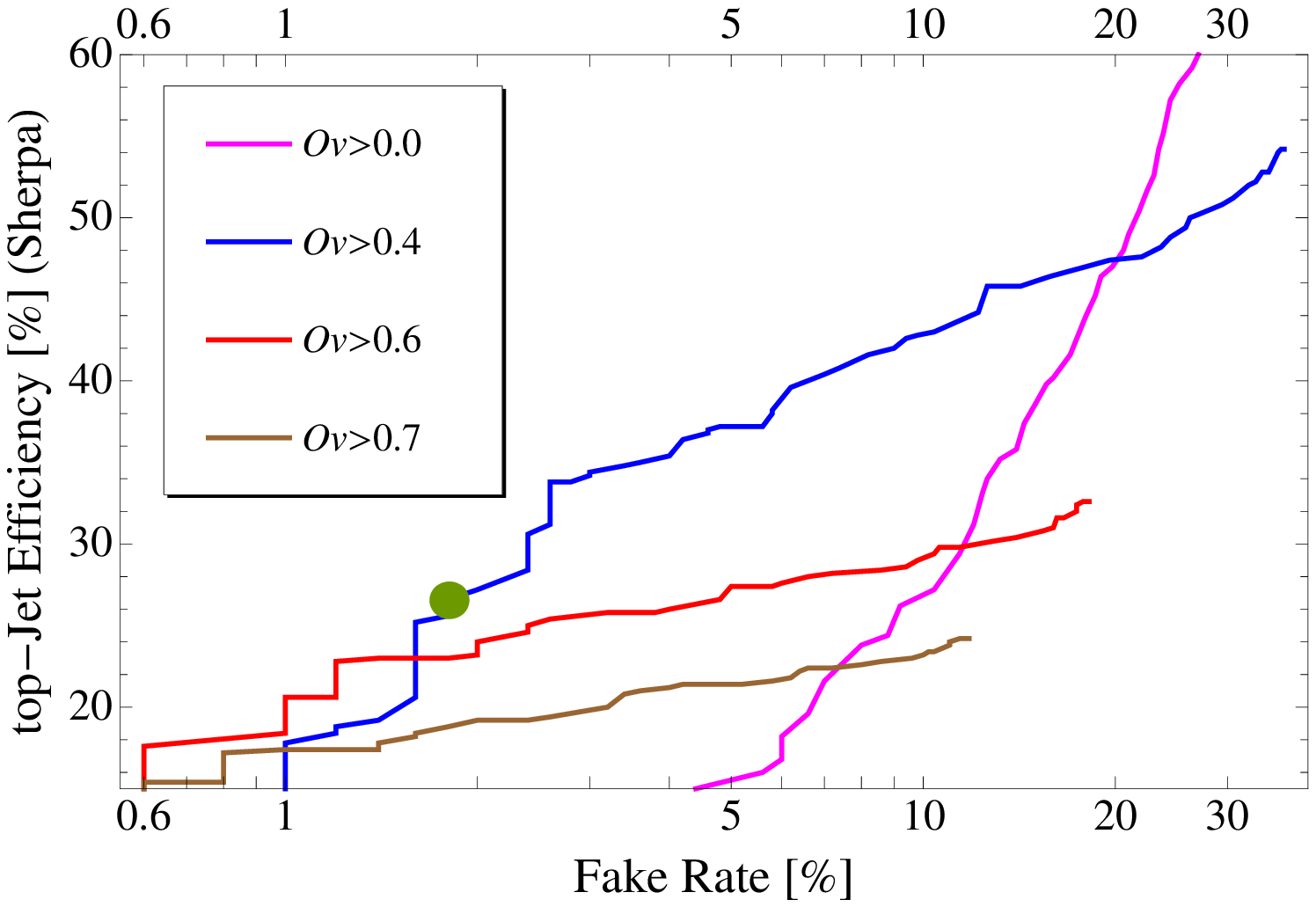}   
\end{tabular}
\caption{
Comparison of fake rate {\it vs.}\ efficiency with various cuts on template overlap $Ov$ and $Pf$ with top jets and QCD jets from different MC [upper left (right) Pythia (MG/ME) and Sherpa on the bottom], for $R=0.5$,  950 GeV$\le P_{0} \le$1050 GeV, 160 GeV$\le m_J \le$190 GeV and  $m_{top}=174$ GeV.  
The lines show the effect of cuts in planar flow ($Pf$) for fixed overlap ($Ov$), with the lowest
(most inclusive) $Pf$ cuts to the right. The green dot is for $Pf>0.6$ and $Ov>0.4$.
}\label{top_probe_function_pf_ratio_comparison}
\end{figure} 

As a simple application of these ideas,
in Fig.~\ref{top_probe_function_pf_ratio_comparison}, we show fake rate {\it vs.}\ efficiency with various cuts on 
the template overlap $Ov$
 as found from Fig.\ \ref{top_probe_function_pf_comparison}.
 For a given cut on $Ov$ denoted by a same-colored line, the efficiency is controlled by the upper cut on $Pf$. 
 Each point on one of these curves corresponds to a specific choice of $Pf$ at fixed $Ov$, and hence to
 the set of points within a rectangle that includes the upper right corners of the corresponding 
 scatter plots in Fig.\ \ref{top_probe_function_pf_comparison}.
 The results depend on the choice of $Ov$ cut, but it is clear that any cut above 0.2 leads to a 
 substantial increase in efficiency.   We present these results for demonstration purposes only,
 and have not carried out a systematic study of how to maximize rejection power.
 
  Our final results for the top jet case are summarized in Table~\ref{tab:jetMass} for the three different event generators, chosen for the best working point found by these simple, naive one-dimensional cuts in $Ov$ and $Pf$.
 It is evident from the numbers presented that the template overlap method works well for events generated by any of the MC generators.
In each case, we find a large
enhancement of signal compared to background, typically of the order of fifteen or more.
Taking into account the rejection of QCD jets by imposing a mass window, these numbers (for a single massive jet) are multiplied by
factors of ten to twenty.
 The template-based approach thus  yields numbers that compare favorably with 
 those found from other methods in the literature (see for example table 9 of Ref.~\cite{Salam:2009jx}). In addition, it allows for systematic improvement, for example by incorporating the effect of gluon emission
 in the template, or by weighting phase space by squared matrix elements.   Because the template method naturally provides scatter plots like
 those in Fig.\ \ref{top_probe_function_pf_comparison}, we can  imagine optimizing cuts on the data.
 We may also investigate improvements in the overlap functional Eq.\ (\ref{3particletemplate}).
 
 Finally, we note that it is evident both from the scatter plots 
 in Fig.\ \ref{top_probe_function_pf_comparison} and the efficiency  distributions 
 in Fig.\ \ref{top_probe_function_pf_ratio_comparison} that the different generators 
 tend to yield different energy flow patterns.   In particular, the green dots on each of the three plots are the result of identical cuts over $Ov$ and $Pf$. This was also noted earlier in the context of the jet mass distribution~\cite{Almeida:2008tp}.
This observation should serve as a caution regarding the interpretation
of tests for all methods, especially those that rely heavily on the
anticipated structure of soft radiation in final states.    

\begin{table}[t]
\begin{center}
\begin{tabular}{c |c c|c c} \hline\hline
\rule{0pt}{1.2em}
MC &  \multicolumn{2}{c|}{Jet mass cut only} &
\multicolumn{2}{c}{ Mass cut + $Ov$ +$Pf$}  \cr
&   Top-jet efficiency~[\%] & fake rate~[\%] &  Top-jet efficiency~[\%] & fake rate~[\%] \cr
\hline Pythia8  &~$58$& $3.6$ &$21$& $0.022$
\\
MG/ME &~$52$& $3.7$& $11$& $0.017$
\\
Sherpa &~$34$& $3.2$& $7$& $0.032$
\\
\hline\hline
\end{tabular}
\caption{\label{tab:jetMass} 
Efficiencies and fake rates for jets  with $R=0.5$ (using anti-$k_T$: $D=0.5$), 950 GeV$\le P_{0} \le$1050 GeV, 160 GeV$\le m_J \le$190 GeV and  $m_{top}=174$ GeV.  
 The left pair of columns shows efficiencies and fake rates found by imposing the jet mass window only.   The right
 pair takes into account the effects of cuts in $Ov$ and $Pf$ in addition to the mass window.
 For the different MC simulations,
we have imposed various cuts on $Ov$ and $Pf$ variables: for Pythia8 $Ov\ge 0.6$ and $Pf\ge 0.4$,
for MG/ME $Ov\ge 0.7$ and $Pf\ge 0.39$ and for Sherpa $Ov\ge 0.6$ and $Pf\ge 0.48$.
}
\end{center}
\end{table}

\section{Two-particle Templates and Higgs Decay}\label{2point}

We now apply the template overlap method to boosted Higgs boson decays.
The following discussion applies as well to electroweak bosons,
because spin produces relatively small effects in the energy flow~\cite{Almeida:2008yp}.
We define the leading order templates in terms of the lowest-order decays of the Higgs,
schematically,
\beeq
| f \rangle =
| h\rangle^{\rm (LO)} = | p_1,p_2\rangle
\,.\eeeq
As above, our template will be
a set of discretized partonic states corresponding to given angular configurations.

The task of disentangling a Higgs signal from a
QCD background is actually more challenging than for the top, because at lowest
order both boosted Higgs and QCD jets consist of two particles.   
Nevertheless, looking only at the information given from the calorimeter,  we can still obtain the measured energy distribution, 
$dE(j)/d\Omega$, and compare it to templates by an overlap function analogous to Eq.\ (\ref{overlap2}).
At lowest order, signal phase space for the Higgs is characterized by particularly simple
kinematic parameters.   For example, in boosted two-particle Higgs
decays, $h\rightarrow b\bar b$, we can characterize the 
final state at fixed $P_0$ by the angle, $\theta_s$ between
the (two-particle) jet axis and the softer of the two particles.
At fixed $z= m_J/P_0 \ll 1$, the distribution in $\theta_s$ is
given by a ``jet" function 
\cite{Almeida:2008yp},
\be
{d J^{h}\over d\theta_s}\propto {1\over \theta_s^3}\, ,
\label{theta_s_higgs}
\ee
rather strongly peaked for small $\theta_s \gg z$.
When $\theta_s$ approaches its minimum value, the decays are ``democratic", sharing
the energy of the Higgs nearly evenly between
the pair.   The distribution for lowest-order QCD events
is still peaked, but much less so 
\cite{Almeida:2008yp}, 
\be
{d J^{\rm QCD}\over d\theta_s}\propto {1\over \theta_s}\, .
\label{theta_s_QCD}
\ee
The two-particle phase space parameter $\theta_s$, of course,
is not a physical quantity.   We can, however, parameterize
the two-particle peak template state, $f[j]$ that a physical state, $j$ most
closely resembles, by matching energy flows, as discussed in Sec.\ \ref{general}.
Once we have identified $f[j]$, 
we can assign a value of $\theta_s$, or any other
kinematic parameter of $f[j]$, to the corresponding physical state $j$.    Template
overlaps enable us to make this identification, and therefore
to make selections among data events based on quantifiable criteria.
In what follows, we apply the peak overlap method introduced in Sec.\ \ref{general} to Higgs decay.

\subsection{Higgs templates}

We now describe the procedure for applying the peak template overlap method for Higgs, with a scheme for discretizing the data.

\subsubsection{Discretization of the data with jet mass and energy selection}

Given a set of data, we impose a jet mass window for the Higgs with a specific cone size $R$ and discretize the data with a convenient mass and energy 
range: for our demonstration we choose the jet mass window to be  110 GeV $\le m_J \le$ 130 GeV, with Higgs mass chosen to be 120 GeV, cone size $R=0.4$ and jet energy 950 GeV $\le P_0 \le$ 1050 GeV.  (For a full analysis, one can discretize the data with a certain step of energy, say 100 GeV, since jet energy is an input of our template function.)   This gives us a set of final states $j$.

For any state $j$, we determine the measured (or MC generated) energy distribution, $dE(j)/d\Omega$, in the physical $\theta$-$\phi$ plane with respect to the jet axis for each reconstructed jet, and
we can start discretizing data into a jet-energy configuration.
In our demonstration for the Higgs, we discretize the $\theta$-$\phi$ plane into cells of size $\Delta \theta=0.04$ and $\Delta \phi=0.1$.
Next, we again assemble a table of energies $E(\rm row_m, column_n)$, where $\rm row_m$ and $\rm column_n$ are the row and column number corresponding to the discretized values of $\theta$ and $\phi$.

\subsubsection{Construction of template function}\label{2point_construction}

As for top decay, we generate our templates
$f$ from a set of discretized angles.   For the two-body Higgs decay, two angles define
the two-body state of the daughter particles.   By analogy to the top case, we choose these
as the polar and azimuthal angles in the Higgs rest frame, relative to the boost axis that
links the Higgs rest frame with the lab frame.
A simple Lorentz transformation determines the momenta of the daughter particles
in the lab frame, where they are compared to data.   

Once again, we generate a large
set of template states, so that we are confident of identifying the peak value of overlap.
We discretize the angles 
with a small length of $2\pi/120$.~\footnote{In fact, one can make it as small as one desires, since we will choose only one of them corresponding to ``peak'' template.}
We can now encode the two physical angles in terms of row and column numbers, corresponding to the data discretization scheme. Each template consists of the information ($\rm row_a, column_a$, $E_{\rm a}$) for each of the two daughter particles.    We exclude those templates having polar
angles larger than the cone size $R$.  Also, if desirable, we impose an energy cut on the template, {\it  i.e.}, removing templates with less than five percent of the total energy.

\subsubsection{Two-particle template overlap}\label{2point_convolution}

We are now ready to implement Eq.\ (\ref{overlap2}) for the Higgs, by defining an overlap between templates, $|f\rangle $, and
jet states $|j\rangle$,  $Ov=\langle j | f \rangle$.
Defined as above, our templates each have two cells corresponding to two daughter partons ($q$ and $\bar q$) with their row and column numbers determined by the data discretization scheme.

As for the top, we compute the overlap between data state $j$ and template $f$
from an unweighted sum of all the energy in the total nine cells 
of state $j$ surrounding and including the two occupied cells of template state $f$.
In summary, we define the overlap of a template function with the energy distribution of the data to be
\beeq
Ov(j,f)  &=&  
{ \rm{max}}_{\tau^{(R)}_n}\ 
\exp\left[\, -\ \sum_{a=1}^2 \frac{1}{2\sigma^2_a}\left(  
\,  \sum_{k=i_a-1}^{i_a+1}\sum_{l=j_a-1}^{j_a+1} E{(k,l)}   - E(i_a,j_a)^{(f)}\, \right)^2
   \right]\, ,
 \label{2particletemplate}
 \eeeq
where $E(i_a,j_a)^{(f)}$ is the energy in the template state for  
particle $a$.   If one of the sums extends outside
the jet cone, we set the corresponding energies $E(k,l)$ to zero.
Again, we fix $\sigma_a$ (for the $a$th parton) by that parton's energy,
$\sigma_a=E(i_a,j_a)^{(f)}/2$, as in Eq.\ (\ref{overlap_convolution2}).

\begin{figure}[hptb]
\begin{center}
\begin{tabular}{c}
\includegraphics[width=.48\hsize]{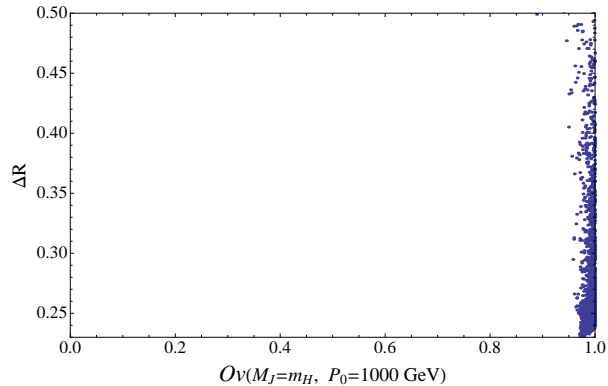} 
\end{tabular}
\caption{A scatter plot of template overlap, Eq.~(\ref{2particletemplate}) and the angular distance $\Delta R$ between the two partons coming from the Higgs decay, with $P_{0}=1$\ TeV, $m_H=120\rm \,GeV$. 
}\label{partonic_level_higgs}
\end{center}
\end{figure}

In Fig.~\ref{partonic_level_higgs}, we 
use the overlap, Eq.\ (\ref{2particletemplate}) to validate the template function 
when compared to MC output events at partonic level, showing that each peak value is close to unity for all the events in our Higgs decaying into a $b\bar b$ sample.
The points cluster even closer to unity for the Higgs than for the top, Fig.\ \ref{partonic_level_top}, because we have used a finer discretization for the (simpler) Higgs templates.

\subsection{Enhancing overlap with $\theta_s$ and angularities}

\begin{figure}[hptb]
\begin{center}
\begin{tabular}{c}
\includegraphics[width=.55\hsize]{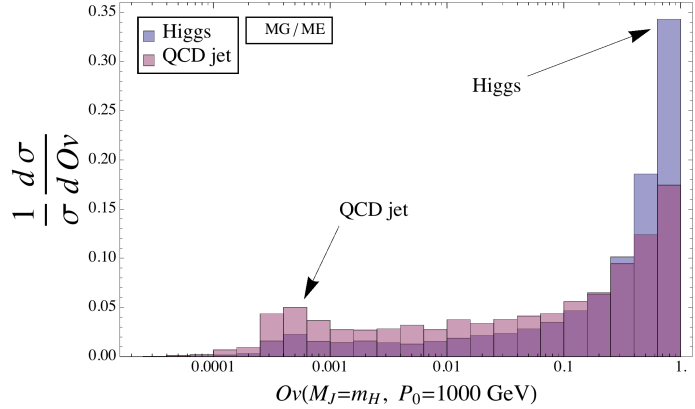} 
\end{tabular}
\caption{Histogram of template overlap distributions (Eq.~(\ref{2particletemplate})) $Ov(j,f_2)$ with MC-generated Higgs jets and QCD jets, where $R=0.4$,  950 GeV$\le P_{0} \le$1050 GeV, 110 GeV$\le m_J \le$130 GeV and  $m_H=120\rm \,GeV$. (MG/ME~\cite{MG} with MLM matching~\cite{MLM}.)
}\label{jet_level_higgs_Probe}
\end{center}
\end{figure} 

We now apply the peak template overlap method to analyze energetic Higgs jet events {\it vis-a-vis} QCD jets.   We use the data
for QCD jet and hadronic Higgs jet events (after showering and hadroniztion), for $R=0.4$, 950 GeV$\le P_{0} \le$1050 GeV, 110 GeV$\le m_J \le$130 GeV and  $m_H=120$ GeV as obtained from MG/ME~\cite{MG} (with MLM matching~\cite{MLM}) via 
anti-$k_T$ jet clustering algorithm~\cite{Cacciari:2008gp}.
 
Our aim is to understand how well we can discriminate our signal from the 
QCD background using the simplest two-particle templates. 
In Fig.~\ref{jet_level_higgs_Probe}, we compare the template overlap $Ov(j,f)$ distributions from Eq.~(\ref{2particletemplate}) for Higgs and QCD jets.
We see that Higgs jet events are peaked toward larger values of template overlap, $Ov$ than QCD jets.
We can therefore use a large $Ov$ value as a quality cut, to ensure that the events under consideration are two-pronged like in terms of the energy flow, say,
$Ov \ge 0.85$.
Furthermore, even within the two-body description, we have seen that Higgs events tend to be peaked towards smaller $\theta_s$ than the QCD jets.   We  hence expect to improve rejection power from an appropriate cut on $\theta_s$.  

To the extent that the energy flow of the jets is similar to that of two-body decay,
their kinematics is determined by a single continuous variable, of which $\theta_s$ 
is only one example.   Indeed, we can use properties of the data itself as alternatives to $\theta_s$.
A set of such alternatives is given the class of angularities, classified by a parameter $a$ and defined by~\cite{Berger:2003iw,Almeida:2008yp}
\begin{eqnarray}
\tilde\tau_a(R,m_J) = \frac{1}{m_J} \sum_{i \in jet} \omega_i\, \sin^a\left(\frac{\pi \theta_i}{2R}\right)\,
\left[ \,1 - \cos\left(\frac{\pi \theta_i}{2R}\right)\, \right]^{1-a}
\sim  
\frac{1}{ m_J}\, \frac{1}{2^{1-a}} \,\sum_{i \in jet} \omega_i\,\left({\pi \theta_i\over 2R }\right)^{2-a}\, ,
\nonumber\\
\label{tauadef}
\end{eqnarray}
where $\omega_i$ is the energy of a component inside the jet (such as a calorimeter tower).
Limiting the parameter $a\leq2$ ensures IR safety, as can be seen from the second expression on the right-hand side of the equation, which is valid for small angle radiation $\theta_i\ll1$.

Angularities, $\tilde \tau_a$, distinguish between Higgs and QCD jets in much the same way as
the template angle $\theta_s$, as can be seen by examining the jet differential distributions in $\tilde \tau_a$,
analogous to Eqs.\ (\ref{theta_s_higgs}) and (\ref{theta_s_QCD}) for $\theta_s$.
In particular, a simple approximation can be obtained when $z=m_J/P_0\ll\theta_s\ll1$ 
and $a$ is negative with $|a|={\cal O}(1)$,
\be
{d J^{h}\over d\tilde\tau_a}\propto {1\over |a| \left(\tilde \tau_a\right)^{1-{2\over a}}}\, ,
\label{tau_higgs}
\ee
rather strongly peaked for small $\tilde \tau_a $.
As suggested above, these decays are ``democratic", sharing
the energy of the Higgs rather evenly between
the pair.   The distribution for lowest-order QCD events
is still peaked at small $\tilde \tau_a$, but less so,
\be
{d J^{\rm QCD}\over d\tilde \tau_a}\propto {1\over |a|\, \tilde \tau_a}\, .
\label{tau_QCD}
\ee
We may thus expect that cuts of the data based on angularities will give results qualitatively
similar to those based on $\theta_s$.
On the other hand, $\theta_s$, which is a parameter for two-body 
template states,  already provides useful information on physical
states, as well as a clear picture of their energy flow.

\begin{figure}[hptb]
\begin{center}
\begin{tabular}{cc}
\includegraphics[width=.48\hsize]{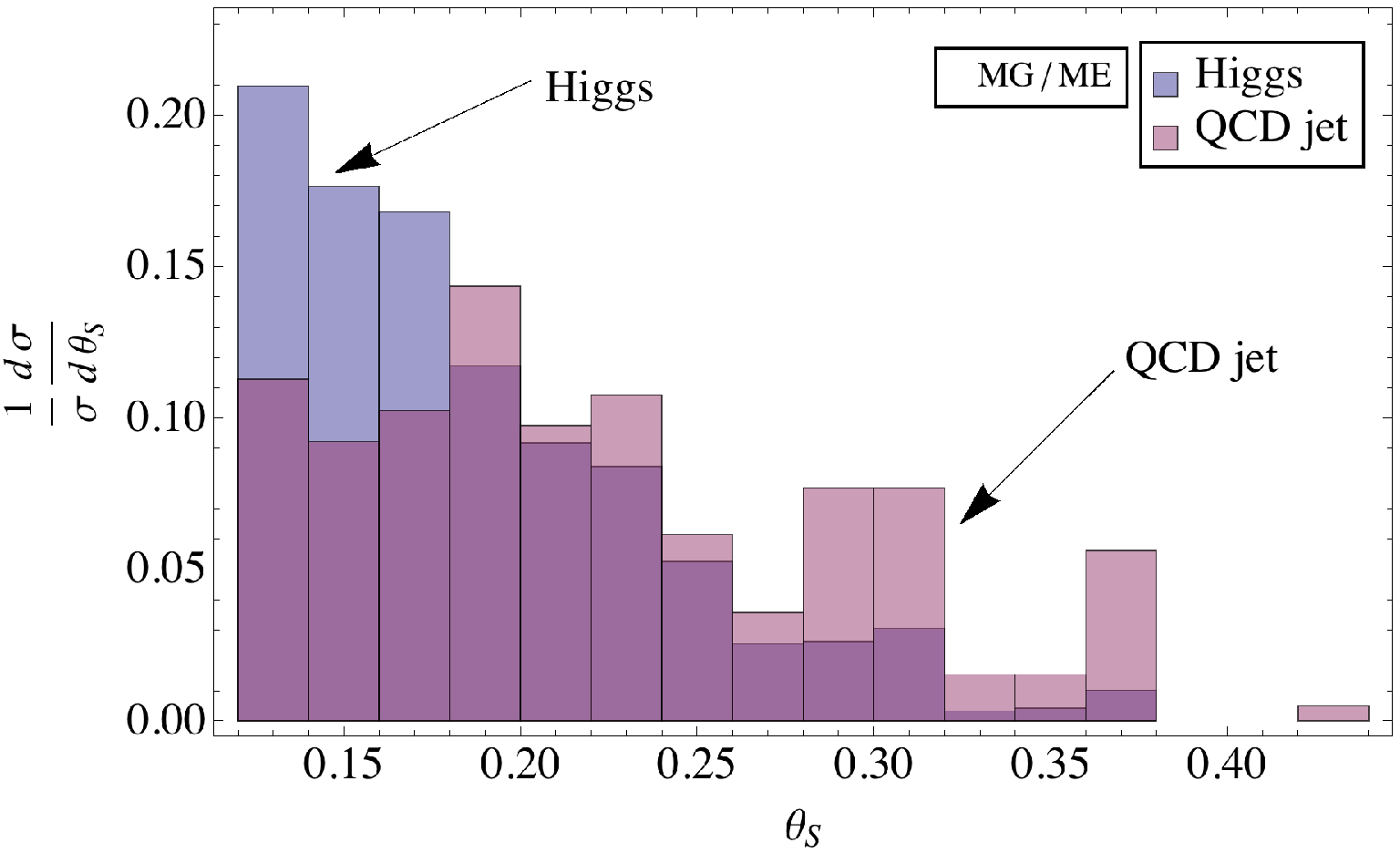}
\includegraphics[width=.48\hsize]{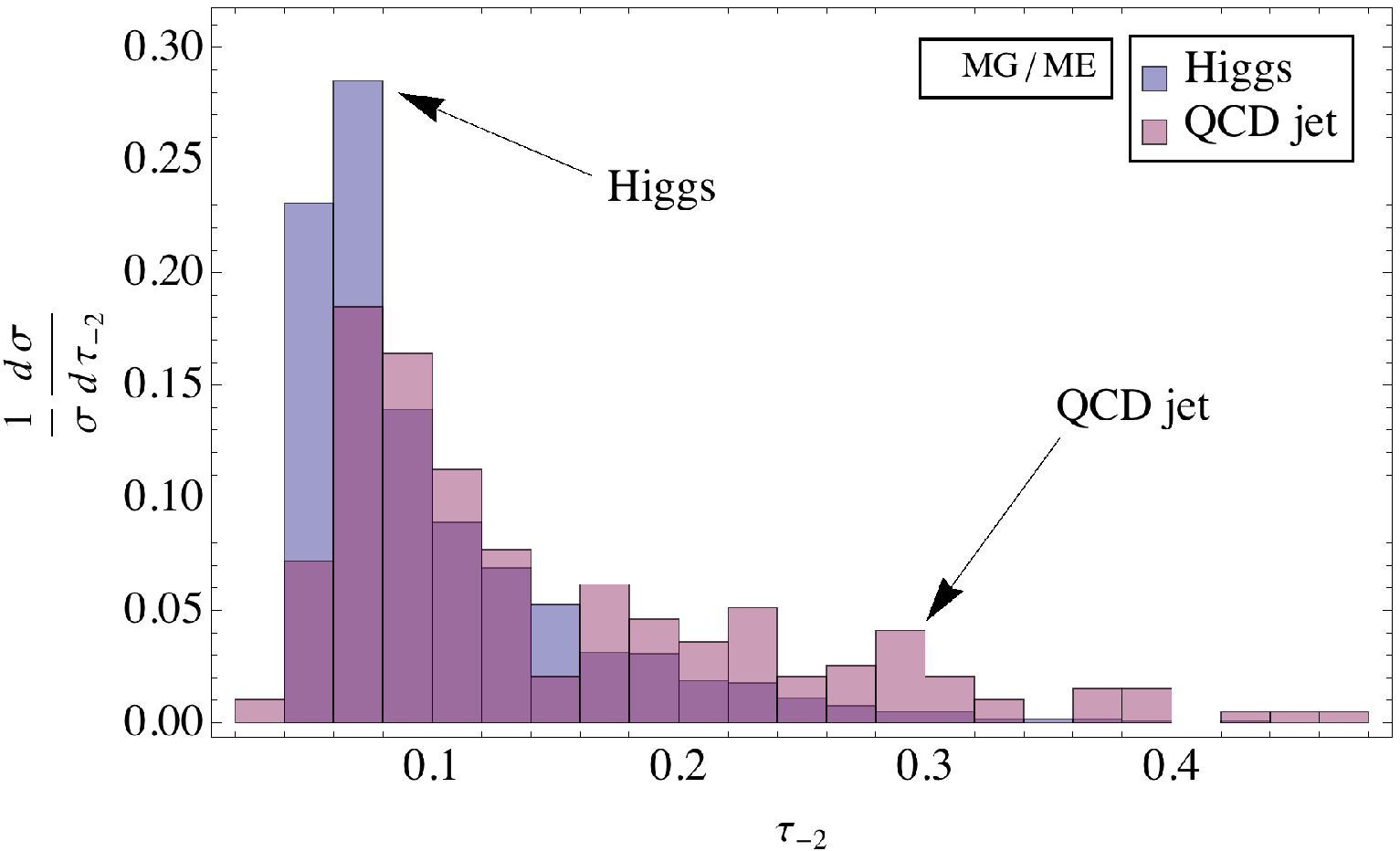} 
\end{tabular}
\caption{
In the plot on the left (right), we show a histogram of $\theta_s$ ($\tilde \tau_{-2}$) with template overlap $Ov\ge$ 0.85.
We choose $R=0.4$,  950 GeV$\le P_{0} \le$1050 GeV, 110 GeV$\le m_J \le$130 GeV and  $m_H=120$ GeV.
(MG/ME~\cite{MG} with MLM matching~\cite{MLM}.)
}\label{jet_level_higgs_Probe_ThetaS}
\end{center}
\end{figure}

\begin{figure}[hptb]
\begin{center}
\begin{tabular}{cc}
\includegraphics[width=.55\hsize]{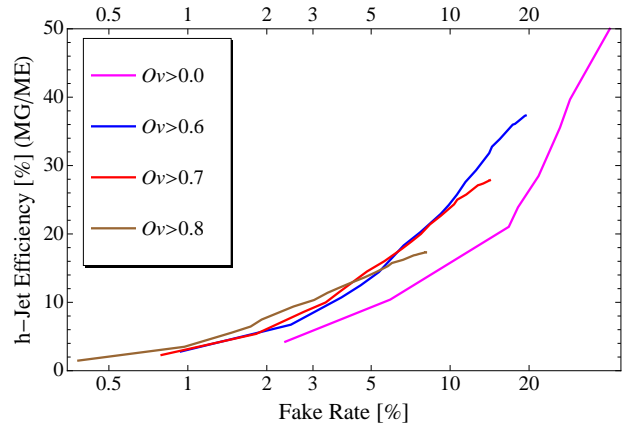}
\end{tabular}
\caption{Fake rate {\it vs.}\ efficiency with various cuts on template overlap $Ov$ and $\theta_s$, for $R=0.4$,  950 GeV$\le P_{0} \le$1050 GeV, 110 GeV$\le m_J \le$130 GeV and  $m_H=120$ GeV. The curves are the result of varying the maximal value of $\theta_s$. Both efficiency and fake rates decrease as we lower the cut on $\theta_s$.
(MG/ME~\cite{MG} with MLM matching~\cite{MLM}.)
}\label{jet_level_higgs_efficiency_rejection}
\end{center}
\end{figure}

We now analyze the effects of limiting  the data to small $\theta_s$ or small angularity. 
In the plot on the left of Fig.~\ref{jet_level_higgs_Probe_ThetaS}, we compare the $\theta_s$ distributions for Higgs and QCD jets, with a lower cut of template overlap $Ov \ge 0.85$,
which confirms our understanding from Eqs.~(\ref{theta_s_higgs}) and~(\ref{theta_s_QCD}). A cut $\theta_s\leq0.2$ (or a corresponding cut on angularity) clearly removes a larger proportion of QCD jets than Higgs jets.
In Fig.~\ref{jet_level_higgs_efficiency_rejection}, we show efficiency {\it vs.}\ fake rates with various cuts on 
template overlap $Ov$. The curves correspond to a variation of the maximum size of the $\theta_s$ template parameter cut.   Each is a scan from 
$\theta_s^{min}$ (which is fixed by the kinematics) to 
$\theta_s^{\rm max}\le0.43$. For a given cut on $Ov$,
the efficiency is controlled by the $\theta_s^{\rm max}$ variable.

\subsection{Planar flow for the Higgs}

So far, we have analyzed Higgs jets
using only template overlaps based on LO partonic decay kinematics.
In principle, the templates can be systematically improved by including 
the effects of gluon emissions, which contain color flow information~\cite{Sung:2009iq,Gallicchio:2010sw}.
Actually, the effects of higher-order effects can be partly captured by
using planar flow \cite{Almeida:2008yp}, which we have already introduced for the top, and defined in Eq.\ (\ref{Pfdef}).
We expect soft radiation from the boosted color singlet Higgs to be concentrated between the $b$ and $\bar b$ decay products.   This is to be contrasted to a jet initiated by a light parton, whose color is correlated with particles in other parts of phase space, producing radiation in the gaps between those particles and the jet system.
Therefore, we expect that planar flow for Higgs jets will be peaked toward a lower value than that of QCD jets.

\begin{figure}[hptb]
\begin{center}
\begin{tabular}{cc}
\includegraphics[width=.48\hsize]{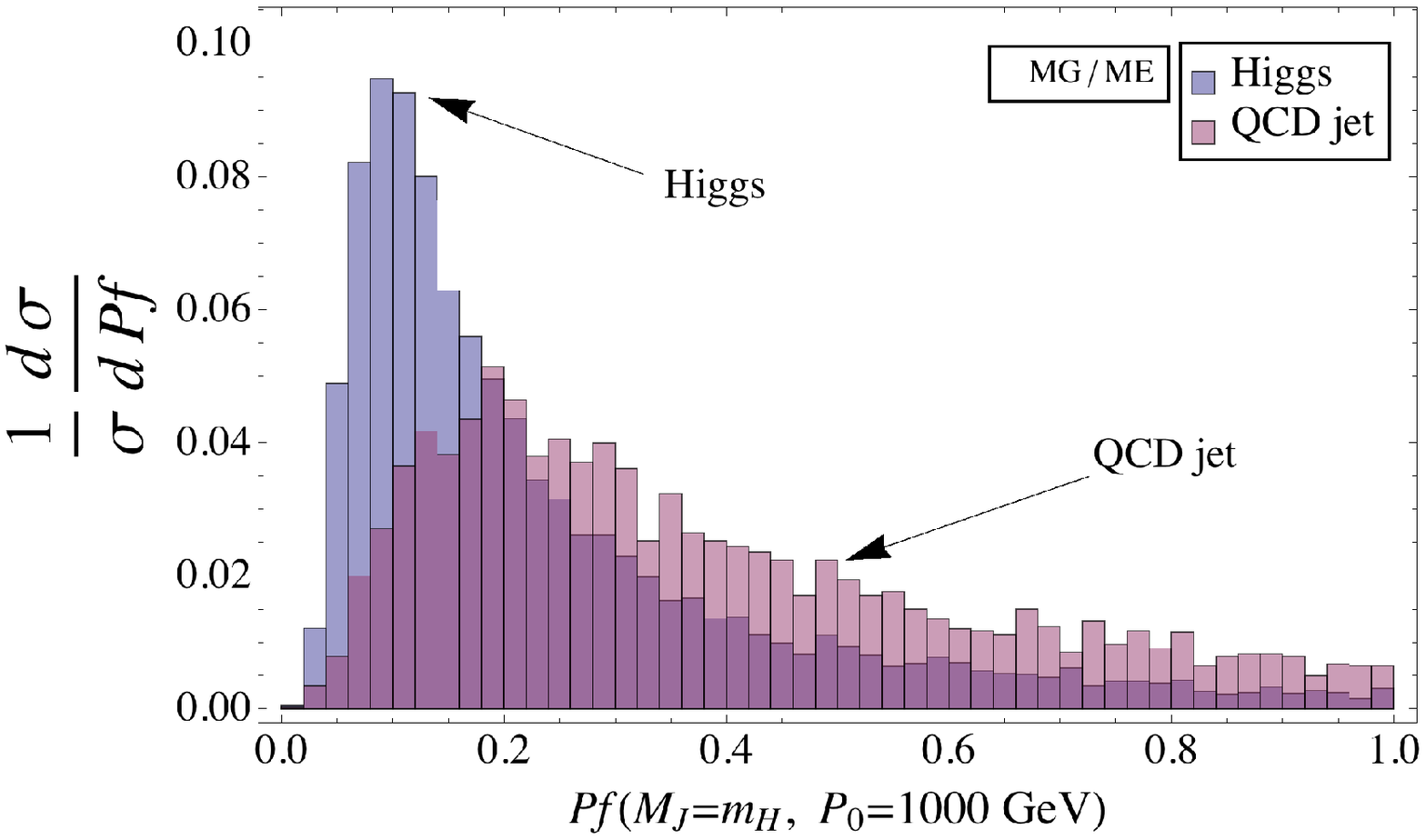}
\includegraphics[width=.48\hsize]{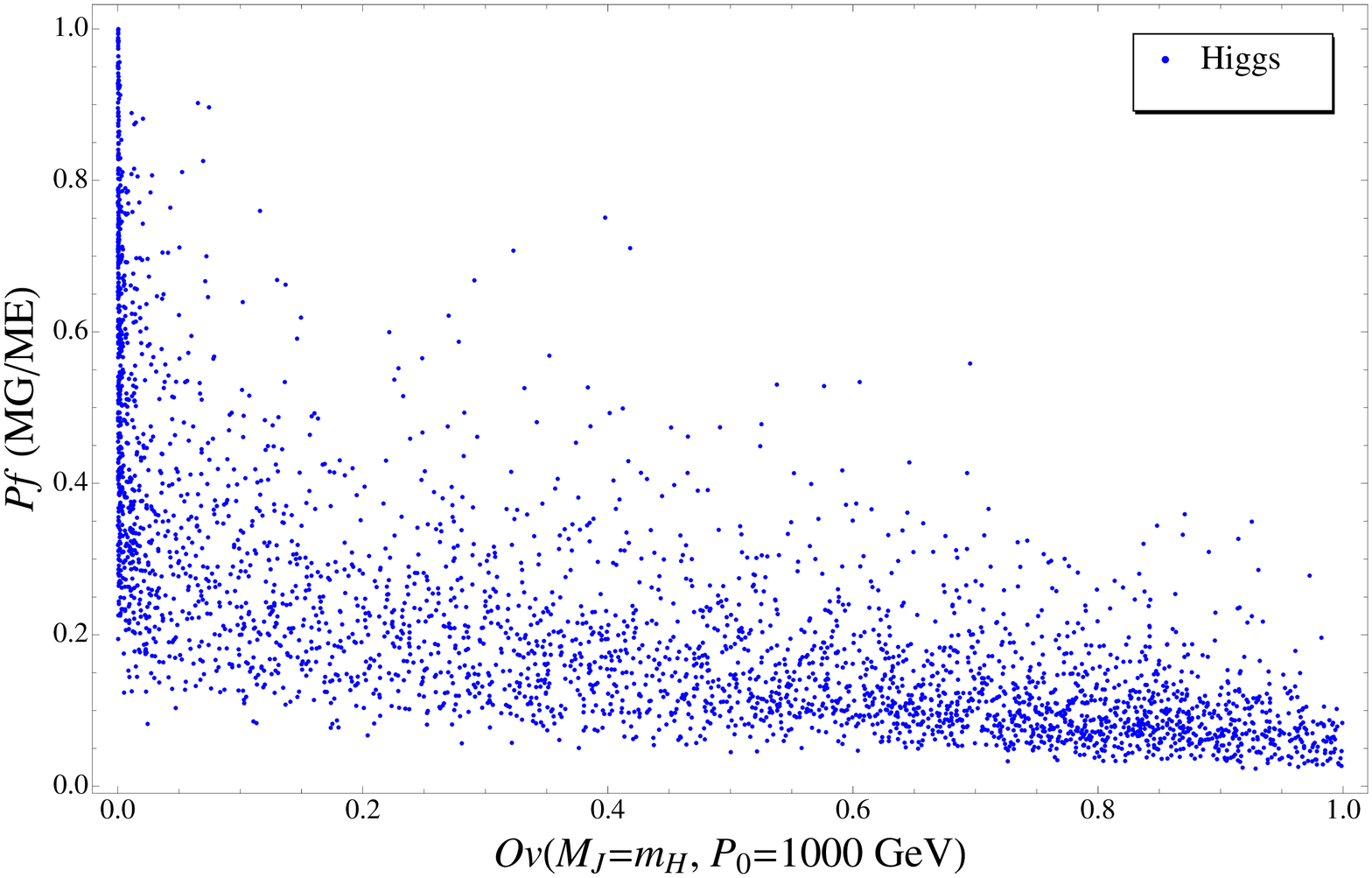}\\
\includegraphics[width=.48\hsize]{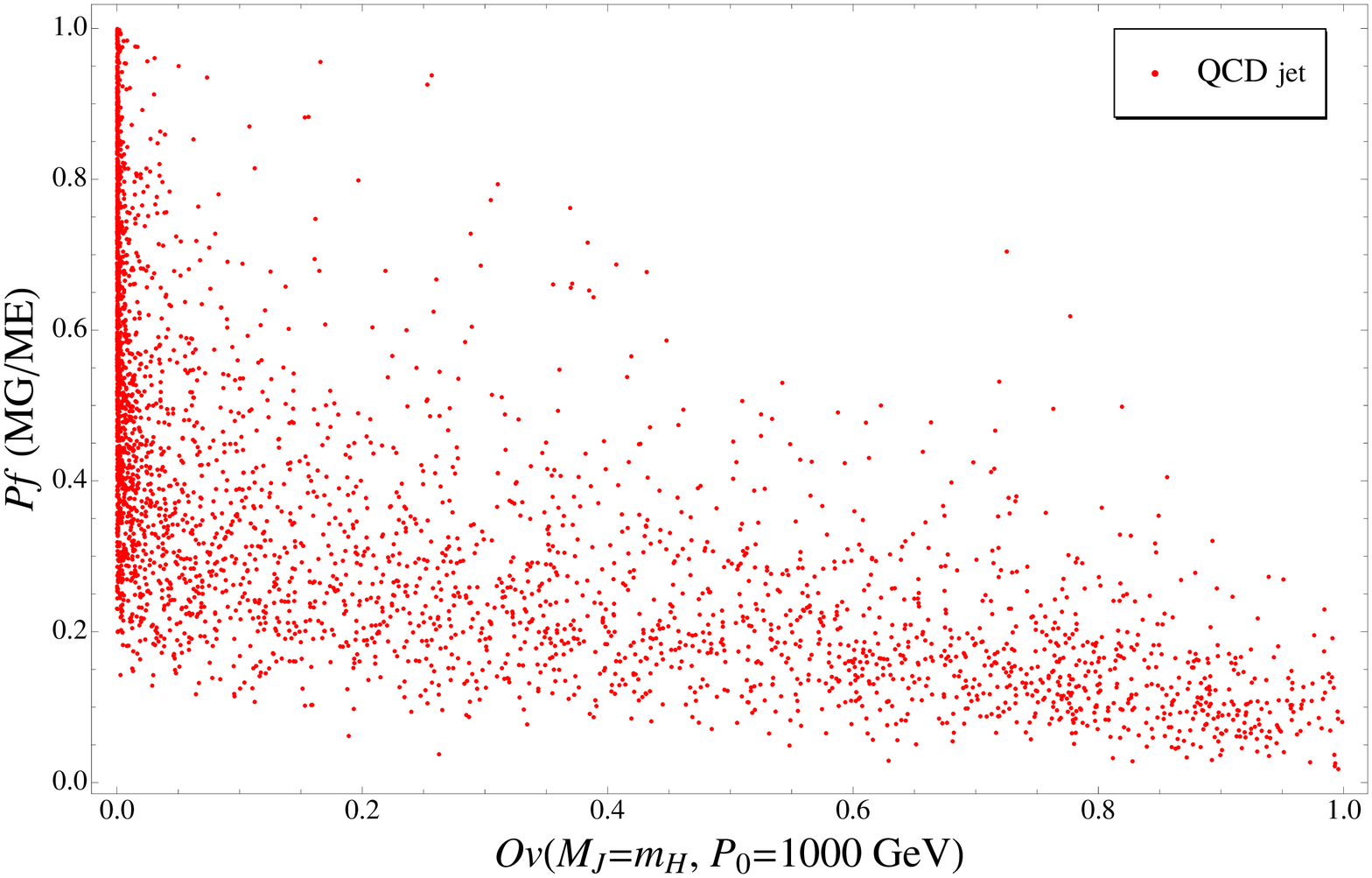}
\end{tabular}
\end{center}
\caption{In the plot on the upper left, we show a histogram of $Pf$ for Higgs jets and QCD jets. In the plot on the upper right, we show a scatter plot of $Pf$ {\it vs.}\ template overlap $Ov$ for Higgs jets.   The remaining plot shows a scattering plot for QCD jets.   Note the  concentration of points for Higgs jets at larger
values of $Ov$ compared to QCD jets.
 We choose $R=0.4$, 950 GeV$\le P_{0} \le$1050 GeV, 110 GeV$\le m_J \le$130 GeV and  $m_H=120$ GeV.
 (MG/ME~\cite{MG} with MLM matching~\cite{MLM})
}\label{jet_level_higgs_pf}
\end{figure}

In Fig.~\ref{jet_level_higgs_pf}, on the upper left, we show the $Pf$ distributions
for QCD jet and Higgs jet events. This panel of the figure confirms our expectation that Higgs jets tend to have smaller $Pf$ values than QCD jets events (for the same $z=m_J/P_0$). 
In the remaining panels, we show  scatter plots of $Pf$ {\it vs.}\ template overlap $Ov$, which show that both QCD and Higgs jets reflect two-pronged energy flow.   In both cases,
those events with large values of $Ov$ tend to have relatively small values of $Pf$.  We see, however, that the Higgs events yield somewhat smaller $Pf$, with a concentration of points at larger $Ov$ in general,  again in agreement with our heuristic expectations.

\begin{figure}[hptb]
\begin{tabular}{cc}
\includegraphics[width=.48\hsize]{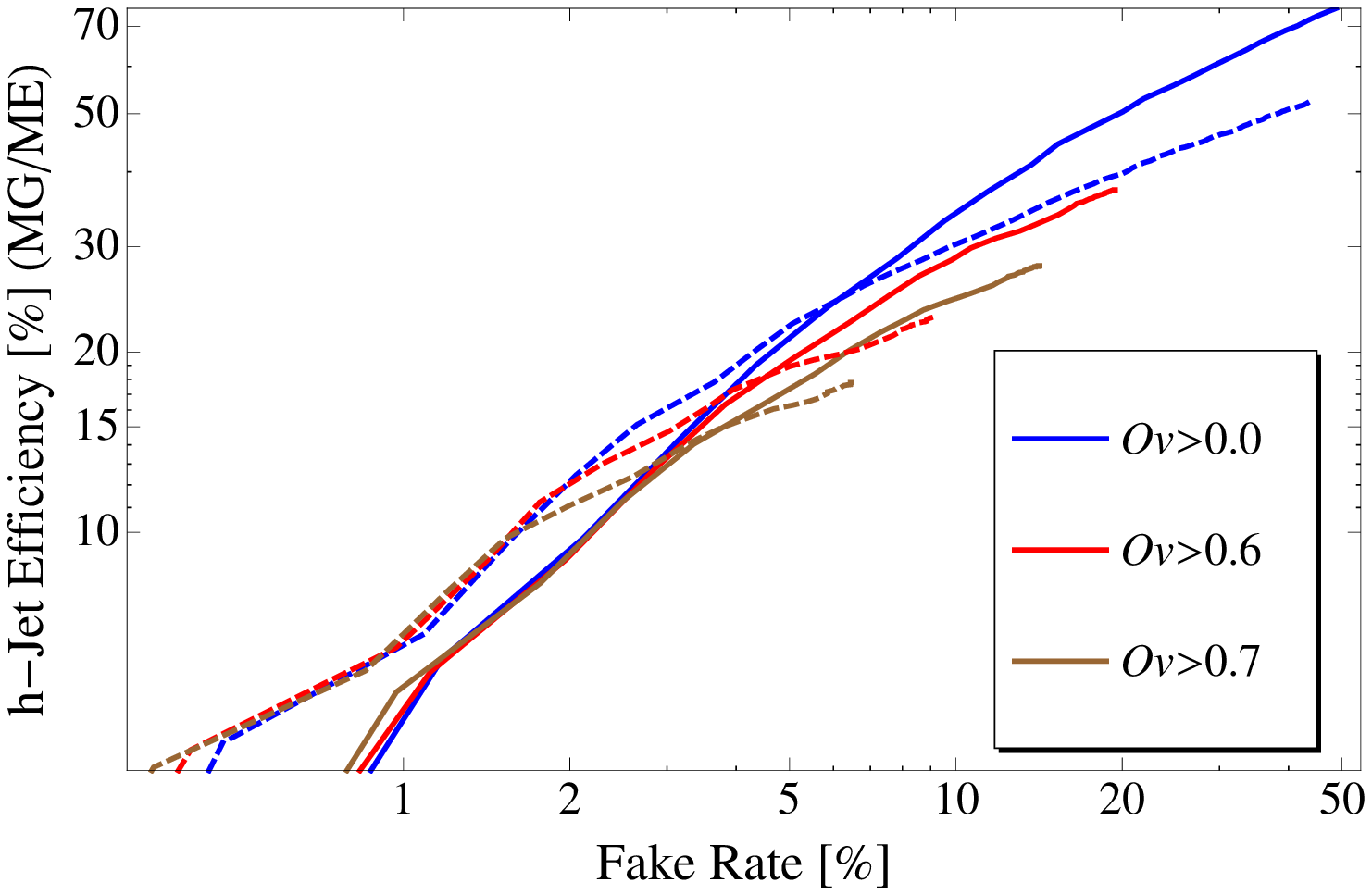}
\includegraphics[width=.48\hsize]{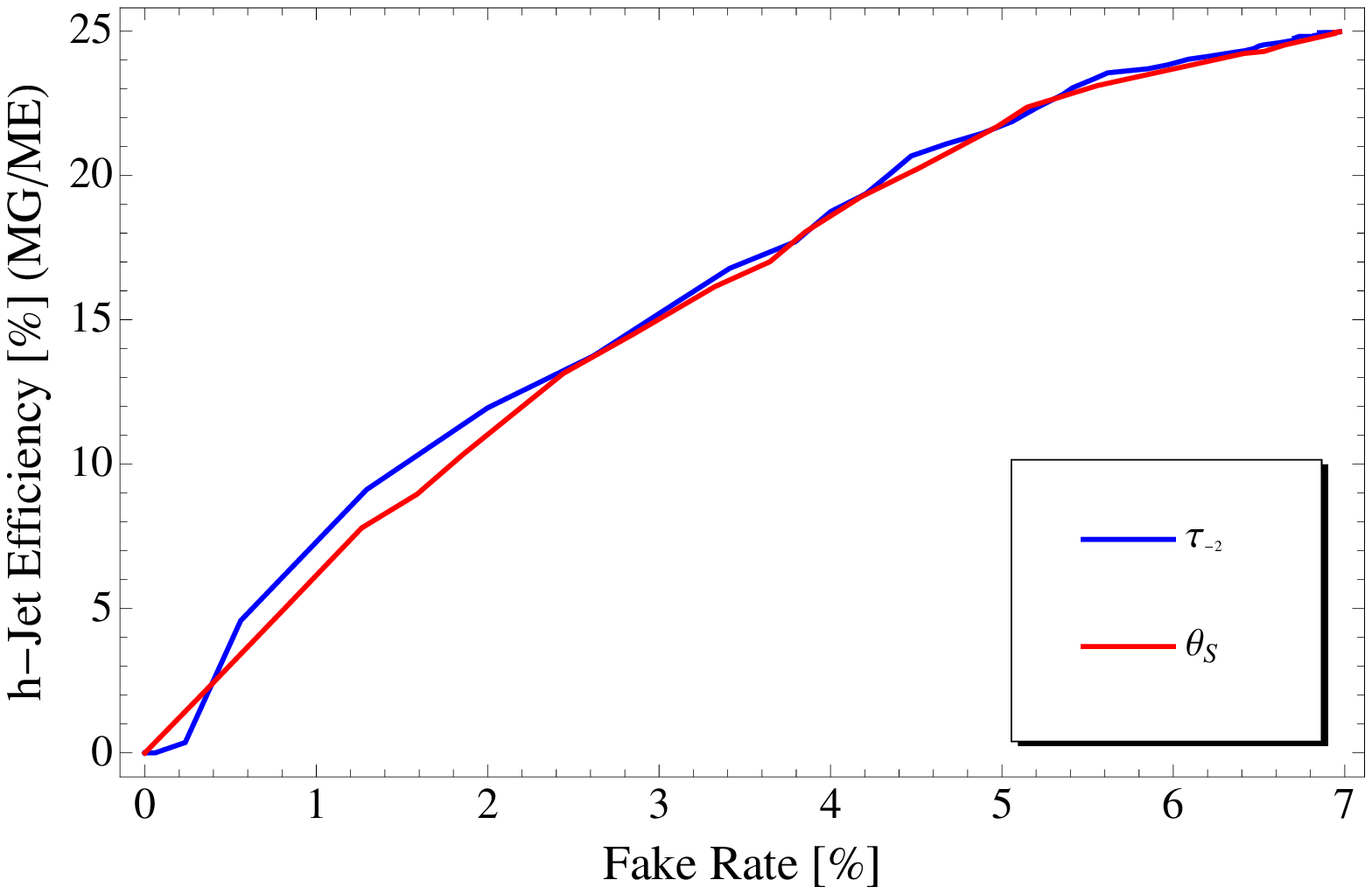}   
\end{tabular}
\caption{
On the left (obtained via MG/ME~\cite{MG} with MLM matching~\cite{MLM}), we show fake rate {\it vs.}\ efficiency, with various cuts of templates $Ov$, while varying the the value of $Pf$ cut, corresponding to the change in efficiency. The dashed lines denote the case when $\theta_s\le 0.2$ cut is implemented, while the solid lines have no $\theta_s$ cut.
In the plot on the right, we show fake rate {\it vs.}\ efficiency with $Pf\le 0.11$ and template overlap cut, $Ov\ge 0.1$,
while varying the value of $\theta_s$ or angularity $\tilde \tau_{-2}$ cut, corresponding to the change in efficiency.
We choose $R=0.4$, 950 GeV$\le P_{0} \le$1050 GeV, 110 GeV$\le m_J \le$130 GeV and  $m_H=120$ GeV.
}\label{jet_level_higgs_pf_ratio}
\end{figure}

Finally, in Fig.~\ref{jet_level_higgs_pf_ratio}, we show the fake rate {\it vs.}\ efficiency when we combine template overlap, $\theta_s$, and planar flow. In the plot on the right, we also show that angularities and $\theta_s$ indeed have similar rejection powers.
Once we combine the fake rate and efficiency 
from   a jet mass cut
(fake rate: 4.5\%, efficiency: 79\%) with template overlap, $\theta_s$, and planar flow, we find, 
for example, 
at efficiency of 9.3\%, a fake rate of 0.084\% (with $Ov \ge 0.5$, $Pf \le 0.09 $, and $\theta_s \le 0.2$).

Once again we point out that rejection power can be expected to improve once the template overlap is extended to  take into account gluon emission. Another interesting but speculative aspect of our method is that, in principle, we can use the LEP data on $Z$ decay and appropriately ``boost" it to match for the relative kinematic regime
to obtain an estimate of the all-orders template from the data itself.

\section{Summary and Conclusions}\label{conclusion}

Template overlaps are a new class of infrared safe jet 
observables, based on functional comparison of the energy
flow in data with the flow in selected sets (the templates) of partonic states.    
We have demonstrated how, even with a relatively naive construction
for the functional, template overlaps 
can be used to enrich samples of highly boosted particle decays 
in the presence of much larger QCD backgrounds.
We have illustrated the method using lowest-order template states for
highly-boosted Higgs and top decays, compared to the outputs of several event generators.
This method, however, relies only on the infrared safety of energy flow, and is more
general than boosted particle decay and may find other applications.
 
 Different event generators give 
 different averages for our template overlaps, which is not surprising since the energy distributions
 within the jets are expected to be sensitive to the showering mechanism, which at present has not been tested experimentally for these kinematical configurations.
 We nevertheless find in each case excellent, although variable, rejection power, defined as the ratio between the signal efficiency and the background fake rate.
 For the Higgs jet case we get a rejection power of order 1:100 and for a single top jet of order
1:1000 (Pythia8), 1:600 (MG/ME), 1:200 (Sherpa) when combined with a jet mass cut, with sizable efficiencies.
The fact that these rejection powers were found to be strong in all cases is encouraging.
 It also suggests that the template overlap method is robust, in the sense that it is
not overly sensitive to the treatment of soft physics.   The latter
clearly varies between the different generators,
which cannot all reproduce the coming 
LHC data.    Differences may be due to 
treatments of multiple interactions, minimum bias and underlaying events as well as showering mechanisms.
The template overlaps described above are capable of systematic improvement by
weighting according to the lowest order matrix elements (in different contexts, such an approach
has been applied to Tevatron data \cite{TeVatronMatrixElement}).
We may also  include higher order
corrections in the template phase space.   
Other improvements may come from changing the functional that defines the overlap,
or from more sophisticated cuts on the data.

{\bf Acknowledgments:} We thank Juan Maldacena for very stimulating discussions. 
We also appreciate the efforts of Steffen Schumann to provide customized Sherpa cut for high $P_T$  QCD and top jet generation.
GP is the Shlomo and Michla Tomarin career development chair; GP is supported by the Israel Science Foundation (grant \#1087/09),
EU-FP7 Marie Curie, IRG fellowship and the Peter \& Patricia Gruber Award.
The work of LA, GS and IS was
supported in part by the National Science Foundation, grants
PHY-0354776, PHY-0354822 and PHY-0653342, and the work of LA 
  was also supported in part by U.S. DOE under contract 
No. DE-AC02-98CH10886.

\end{document}